\documentclass[prx,amsmath,amssymb,twocolumn,showpacs]{revtex4}

\usepackage{graphicx}
\usepackage{subfigure}
\usepackage{adjustbox}
\usepackage{bm}
\usepackage{color}
\usepackage{braket}
\usepackage{standalone}
\usepackage{multirow}
\usepackage{tikz}
\usepackage{mathrsfs}
\usepackage{diagbox}
\usepackage{array}
\usepackage[colorlinks,bookmarks=true,citecolor=blue,linkcolor=red,urlcolor=blue]{hyperref}

\begin{document}
\title{Boundaries of boundaries: A systematic approach to lattice models with solvable boundary states of arbitrary codimension}

\author{Flore K. Kunst$^{1}$, Guido van Miert$^{2}$ and Emil J. Bergholtz$^{1}$}

\affiliation{$^1$ Department of Physics, Stockholm University, AlbaNova University Center, 106 91 Stockholm, Sweden
\\$^2$ Institute for Theoretical Physics, Centre for Extreme Matter and Emergent Phenomena, Utrecht University, Princetonplein 5, 3584 CC Utrecht, The Netherlands}
\date{\today}

\begin{abstract}
We present a generic and systematic approach for constructing $D-$dimensional lattice models with exactly solvable $d-$dimensional boundary states localized to corners, edges, hinges and surfaces. These solvable models represent a class of ``sweet spots" in the space of possible tight-binding models---the exact solutions remain valid for any tight-binding parameters as long as they obey simple locality conditions that are manifest in the underlying lattice structure. Consequently, our models capture the physics of both (higher-order) topological and nontopological phases as well as the transitions between them in a particularly illuminating and transparent manner. 

\end{abstract} 

\pacs{71.10.Fd, 73.21.Ac, 73.20.At, 03.65.Vf}

\maketitle

\section{Introduction}

\noindent Topological phases of matter form one of the central topics in condensed matter physics due to their exotic properties, such as the presence of robust boundary states \cite{hasankane, qizhang, Weylreview}. The most prominent manifestation of these phases comes in the form of topological insulators, which have a gapped $D$-dimensional bulk spectrum while supporting a dissipationless current on their $(d = D-1)$-dimensional boundaries. Celebrated examples of these modes are the chiral, edge states of quantum Hall and Chern insulators \cite{klitzingdordapepper, haldane, tknn, hofstadter, changzhangfengshenzhang, jotzumesserdesbuquoislebratuehlingergreifesslinger}, the helical, edge states of two-dimensional $\mathbb{Z}_2$ insulators \cite{kaneandmele, kaneandmeletwo, bernevigzhang, bernevighugheszhang, koenigwiedmannbruenerothbuhmannmolenkamp}, and the Dirac surface states of three-dimensional strong topological insulators \cite{kanemelestrongti}. Closely related to these models are topological semimetals whose bulk is instead gapless but nevertheless host robust states on their boundaries. A paradigmatic example of this phase is the Weyl semimetal, which features Weyl nodes in its low-energy description, and supports the existence of Fermi arcs on its surfaces \cite{volovik, murakami, wanturnerbishwanathsavrasov, burkovbalents, luwangyeranfujoannopoulossoljacic, xubelopolskialidoustetal, lvwengwantmiaoetal}. 

Recently, an extended family of topological models were proposed in the form of higher-order topological insulators \cite{sittefoschaltmanfritz, benalcazarbernevighughes, langhehnpentrifuoppenbrouwer, linhughes, songfangfang, benalcazarbernevighughesagain, schindlercookvergnio, geiertrifunovichoskambrouwer, trifunovicbrouwer, vanmiertortix, kooivanmiertortix}. In this case, in-gap modes localize to boundaries with codimension $(D-d)>1$, e.g., the zero-dimensional corners of two- or three-dimensional latices, or the one-dimensional hinges of three-dimensional lattices. The existence of these $(D-d)$th-order modes can be guaranteed in the presence of crystalline symmetries that relate the edges or surfaces adjacent to higher order boundaries to one another such that the corner or hinge connecting them form true phase boundaries \cite{geiertrifunovichoskambrouwer}.

While solutions for first-order topological boundary states have been derived in several specific models \cite{kitaev,liuqizhang,maokuramotoimurayamakage, shenshanlu, koenigbuhmannmolenkamphughesliuqizhang, mongshivamoggi, zhouluchushenniu,ojanen, aklt, mahyaehardonne, mahyaehardonne2} and several general approaches have been developed to retrieve them \cite{transfer3, alasecobaneraortizviola, duncanoehbergvaliente, cobaneraalaseortizviola}, there is a surprising lack of methods to find analytical solutions for these boundary wave functions that are straightforward and transparent and can be used to describe modes of any codimension. Such a method is not only of theoretical relevance but also provides practical insight on how to engineer lattices that support these states. Indeed, the corner modes predicted to exist in breathing kagome lattices \cite{kunstvmiertbergholtz, ezawapap, xuxuewan}, as discussed in Sec.~\ref{secboundcodimtwo}, have recently been experimentally observed \cite{xueyanggaochongzhang, niweineraluetal, elhassankunstmoritzandlerbergholtzbourenanne}.

\begin{figure}[t]
{\includegraphics[width=\linewidth]{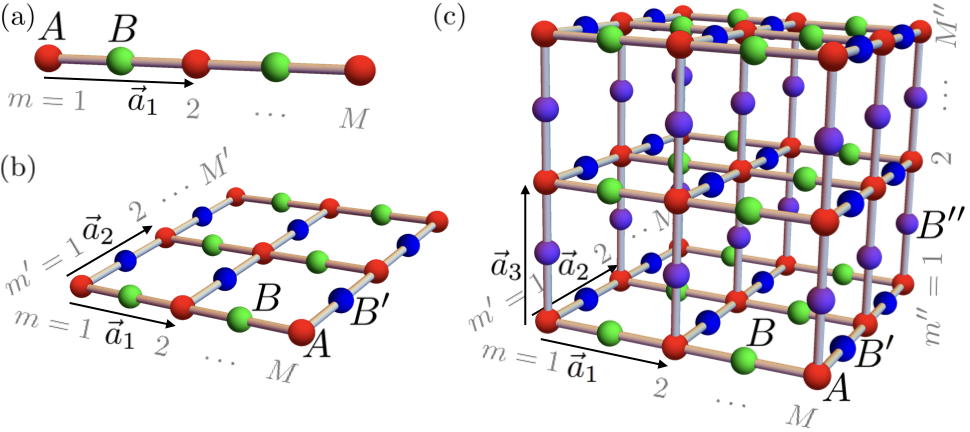}}
\caption{Schematic depiction of lattice structures with $A$, $B$, $B'$, and $B''$ sublattices in red, green, blue, and purple, respectively. The $A$ sublattices are labeled with $m^{(s)}$ and the lattice vectors $\vec{a}_i$ are explicitly indicated. The boundaries of the lattices have codimensions (a) one, (b) two, and (c) three.}
\label{figschemlat}
\end{figure}

In this paper, building on the brief list of specific examples given in our recent publication \cite{kunstvmiertbergholtz}, we give a detailed account of a generic method on how to construct such lattice models, which fulfills the aforementioned criteria by allowing for the engineering of the boundary dispersion, giving access to exact solutions for the boundary-mode wave functions and providing the tools to choose the desired localization of these modes. We make explicit use of exact, destructive interference, which is naturally present on a large family of lattices and allows us to write down exact wave-function solutions whose details depend almost solely on a reoccurring structure of the total system, sublattice $A$, and whose eigenvalues stem from the ``on-site" Hamiltonian on $A$. The localization of these modes is dictated by the indirect connection between different $A$ sublattices via $B^{(s)}$ sublattices [see Figs.~\ref{figschemlat} and \ref{figschemlatmir}], which can be readily adjusted to obtain the desired localization. We thus have the power to \emph{choose} the codimension, the localization, and the dispersion of the exact wave-function solution. While our method thus allows us to retrieve wave-function solutions of any codimension in an exact and straightforward manner, it does not give direct information pertaining to the ground state of the system nor the topological protection of these states. Indeed, additional computations are required to make that determination in which case our exact solutions may be of aid, as we indeed showed in Ref.~\onlinecite{kunsttrescherbergholtz}.

This method has already been investigated and proven to be extremely useful to describe first-order topological phases on frustrated lattice models of the type shown in Fig.~\ref{figschemlat}(a) to find the chiral, edge states of Chern insulators and Fermi arcs of Weyl semimetals \cite{kunsttrescherbergholtz,bergholtzliutreschermoessnerudugawa,trescherbergholtz}. In the current paper, this method is extended to a more general family of lattices and to topological phases of any order. In Ref.~\onlinecite{kunstvmiertbergholtz}, we briefly showed how to obtain exact solutions for corner modes of two-dimensional breathing kagome and three-dimensional breathing pyrochlore, as well as hinge states in two different three-dimensional lattices, one formed of stacked honeycomb lattices and one consisting of stacked Rice-Mele chains \cite{kunstvmiertbergholtz}. Here, we discuss these results in great detail, and provide a generic and in-depth description on how to construct these models. We also introduce an additional example in which we realize chiral, hinge states on the pyrochlore lattice by stacking two-dimensional $\mathbb{Z}_2$ insulators and connecting them in such a way that the helical edge states gap out and only chiral modes on the hinges survive. This technique of creating higher order topological phases has recently also been put forward in Ref.~\onlinecite{trifunovicbrouwer}, where a $D$-dimensional model with second-order modes is created by stacking $(D-1)$-dimensional systems with an alternating topological invariant. In contrast, we couple our topologically nontrivial systems to each other via a trivial layer consisting of $B''$ sublattices, which, due to the disappearing amplitude of the wave function, does not actively contribute to the boundary wave function. We have exemplified some of the different $D$-dimensional models with $d$-dimensional solvable boundary states in Table~\ref{tabdiffmodels}.

\begin{table}[t]
    \begin{tabular}{  l || >{\centering\arraybackslash}p{2.3cm} | >{\centering\arraybackslash}p{2.3cm} | >{\centering\arraybackslash}p{2.3cm} |}
    \backslashbox{D}{d} & 0 & 1 & 2 \\ \hline \hline
    1 & SSH chain & - & - \\ \hline
    2 & Breathing kagome & Chern insulators, $\mathbb{Z}_2$ insulator \cite{kunsttrescherbergholtz} & - \\ \hline
    3 & Breathing pyrochlore & Hinge insulators & Strong TI \cite{kunstvmiertbergholtz2}, Weyl semimetals \cite{kunsttrescherbergholtz} \\
    \hline
    \end{tabular}
    \caption{Examples of $D$-dimensional models that host $d$-dimensional boundary modes. Solvable models corresponding to each of these can be constructed using the framework presented in this work. Details are given here or in the indicated references.}
\label{tabdiffmodels}
\end{table}

This paper is ordered as follows. In Sec.~\ref{sectgenmet}, we introduce our method of constructing lattices that have exact wave-function solutions in its full generality. This is followed by a detailed description of the Hamiltonians on each of the individual lattice structures in Figs.~\ref{figschemlat}(a), \ref{figschemlat}(b), \ref{figschemlat}(c), and \ref{figschemlatmir}, an explicit derivation of the exact wave-function solutions, as well as concrete and relevant examples to each of the lattices in Secs.~\ref{seccodimone}-\ref{secmirrorsymm}, respectively. We conclude in Sec.~\ref{secconclusion}.

\section{General method} \label{sectgenmet}

\begin{figure}[t]
{\includegraphics[width=\linewidth]{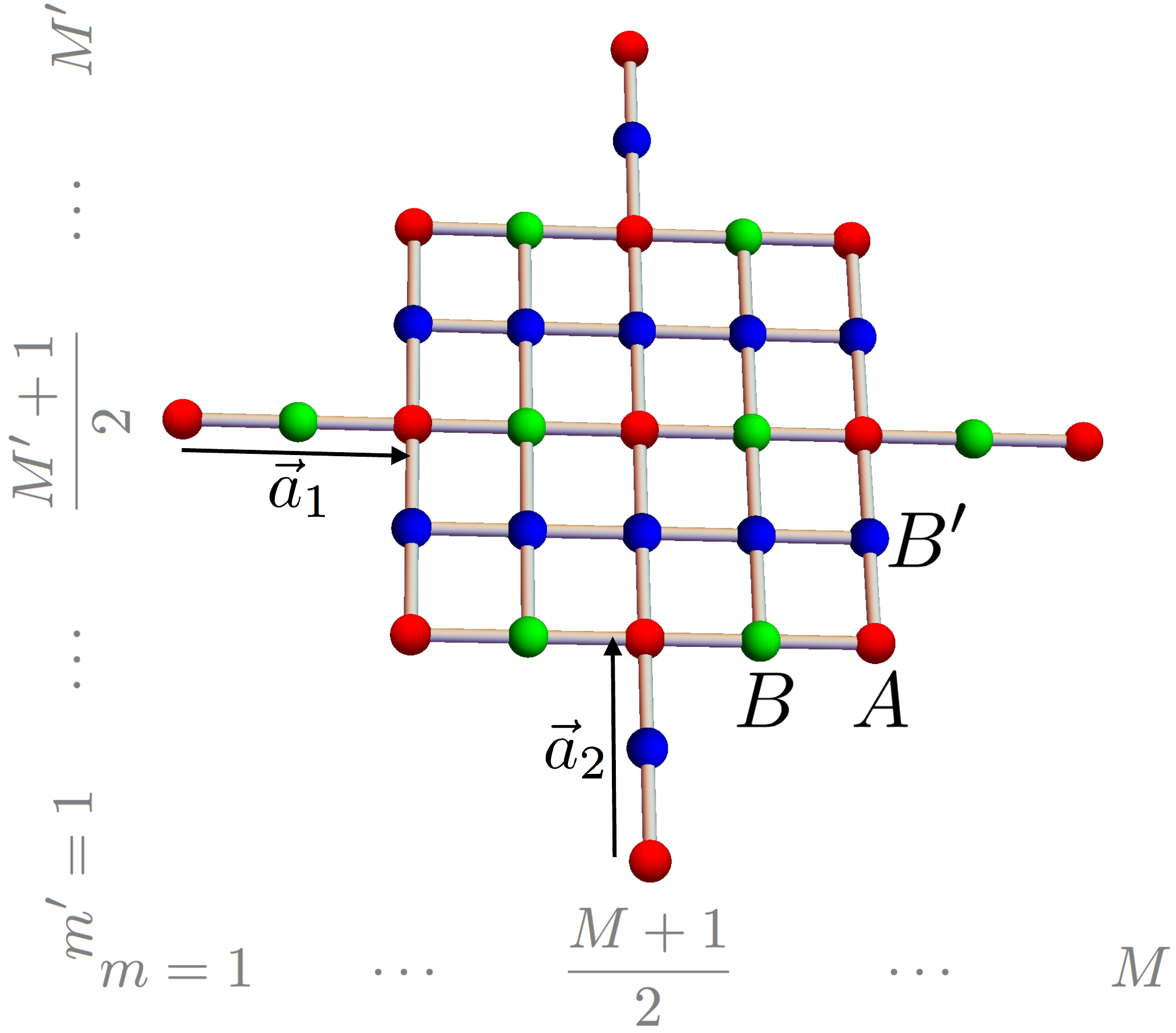}}
\caption{Schematic depiction of a lattice structure with mirror symmetry around $m' = (M'+1)/2$ with the $A$, $B$, and $B'$ sublattices in red, green, and blue, respectively. The $A$ sublattices are labeled with $m^{(s)}$ and the lattice vectors $\vec{a}_i$ are explicitly indicated. The boundaries of this lattice have codimension two.}
\label{figschemlatmir}
\end{figure}

\noindent We start by constructing a general family of lattices whose primary building blocks consist of $A$, $B$, $B'$, $\ldots$, $B^{(D-d-1)}$ sublattices, where the number of sublattices $B^{(s)}$ is directly related to codimension $(D-d)$ of the boundary under investigation. The $D$-dimensional lattices with open boundary conditions in $(D-d)$ dimensions are then formed by connecting the $A$ sublattices with the $B^{(s)}$ sublattices in the directions of the lattice vectors $\vec{a}_{s+1}$ in an alternating fashion such that the boundaries are formed by $A$ sublattices. Figure~\ref{figschemlat} shows a schematic depiction of the simplest examples of lattices constructed in this fashion with boundaries of codimensions one [Fig.~\ref{figschemlat}(a)], two [Fig.~\ref{figschemlat}(b)], and three [Fig.~\ref{figschemlat}(c)], while the schematic lattice in Fig.~\ref{figschemlatmir} with boundaries of codimension two also falls within this class of lattices.

By allowing no direct hoppings between the $A$ sublattices, these lattices naturally support the exact disappearance of the wave-function amplitude of $n$ wave functions on the $B^{(s)}$ sublattices with $n$ the number of degrees of freedom on the $A$ sublattices. The associated eigenvalues $\epsilon_{A_i}$ of these wave functions are the eigenvalues of the Hamiltonian $h_A$, which is the ``on-site" Hamiltonian on each $A$ sublattice. Wave functions of this form read
\begin{align}
&\ket{\psi_i (\vec{k}_{||})} = \mathcal{N}_i (\vec{k}_{||}) \nonumber \\
& \times \sum_{\{m^{(s)}\}} \left\{\prod_{s=0}^{D-d-1} \left[r_i^{(s)} (\vec{k}_{||}) \right]^{m^{(s)}} \right\} c_{A_i, \vec{k}_{||}, \{m^{(s)}\}}^\dagger \ket{0}, \label{eqexactsolgen}
\end{align}
where $\vec{k}_{||}$ is the quasi momentum parallel to the direction of the open boundary conditions, $i = 1, 2, \ldots, n$ labels the solution, $\mathcal{N}_i (\vec{k}_{||})$ is the normalization factor, $m^{(s)}$ labels the $A$ sublattices in the $\vec{a}_{s+1}$ direction with a total of $\sum_s M^{(s)}$ $A$ sublattices, $c_{A_i, \vec{k}_{||}, \{m^{(s)}\}}^\dagger$ creates an electron on sublattice $A$ in unit cell $\{m^{(s)}\}$ with energy $\epsilon_{A_i}$, and $r_i^{(s)} (\vec{k}_{||})$ is obtained by using exact, destructive interference on the $B^{(s)}$ sublattices.

The localization of this wave function on each unit cell is determined by $P_{i, \{m^{(s)}\}} (\vec{k}_{||}) = \braket{\psi_i (\vec{k}_{||})|\Pi_{\{m^{(s)}\}}|\psi_i (\vec{k}_{||})}$, where $\Pi_{\{m^{(s)}\}} = \ket{e_{\{m^{(s)}\}}}\bra{e_{\{m^{(s)}\}}}$ is the projection operator onto the unit cell $\{m^{(s)}\}$, leading to
\begin{equation}
P_{i, \{m^{(s)}\}} (\vec{k}_{||}) = \left|\mathcal{N}_i (\vec{k}_{||})\right|^2 \prod_{s=0}^{D-d-1} |r_i^{(s)} (\vec{k}_{||})|^{2 m^{(s)}},
\end{equation}
which only has nonzero weight on the $A$ sublattices. We thus find that if $|r_i^{(s)}(\vec{k}_{||})|=1, \, \forall s$, the wave function in Eq.~(\ref{eqexactsolgen}) is equally localized on all $A$ sublattices, meaning that it corresponds to a bulk state and the energy band $\epsilon_{A_i}$ necessarily attaches to the bulk bands. However, if $|r_i^{(s)}(\vec{k}_{||})|\neq1$, the mode is exponentially localized to one of the boundaries and the solution in Eq.~(\ref{eqexactsolgen}) describes a boundary mode. More specifically, taking the model in Fig.~\ref{figschemlat}(b) as an example, if $|r_i(\vec{k}_{||})| \neq 1$ and $|r'_i(\vec{k}_{||})| \neq 1$, the wave function in Eq.~(\ref{eqexactsolgen}) localizes to the corners $\{m,m'\} = \{1,1\}$, $\{1,M'\}$, $\{M,1\}$, or $\{M,M\}$ depending on the exact values of $|r_i(\vec{k}_{||})|$ and $|r'_i(\vec{k}_{||})|$. If instead only one of the $|r_i^{(s)}|$ differs from one, e.g., $|r_i(\vec{k}_{||})|=1$ and $|r'_i(\vec{k}_{||})| \neq 1$, the solution in Eq.~(\ref{eqexactsolgen}) localizes to an edge, in this case $\{m,m'\} = \{m, 1\}$ or $\{m, M'\}$ depending on the specific value of $|r'_i(\vec{k}_{||})|$.

Solutions on lattices of the form of Fig.~\ref{figschemlat}(a) are studied in great detail in the context of frustration in Ref.~\onlinecite{kunsttrescherbergholtz} and are generalized in Sec.~\ref{seccodimone}. The boundary modes are captured by Eq.~(\ref{eqexactsolgen}) with $s = 0$ and localize to the $A$ sublattice $m = 1$ ($M$) when $|r_i(\vec{k}_{||})| < 1$ ($>1$). While the schematic lattice exactly corresponds to the famed Su-Schrieffer-Heeger (SSH) model \cite{ssh}, which we indeed use as an example in Sec.~\ref{seccodimone} to illustrate our findings, by dimensional extension it also describes two- and three-dimensional systems with each of the sites now corresponding to one-dimensional, periodic chains and two-dimensional, periodic planes, respectively \cite{kunsttrescherbergholtz}. We can thus use our results to find exact solutions for a plethora of different systems, such as for the chiral edge states of Chern insulators and Fermi arcs of Weyl semimetals \cite{kunsttrescherbergholtz} as well as for the helical edge states of two-dimensional $\mathbb{Z}_2$ insulators and Dirac surface states of three-dimensional strong topological insulators \cite{kunstvmiertbergholtz2} as summarized in Table~\ref{tabdiffmodels}.

The schematic lattices in Figs.~\ref{figschemlat}(b) and \ref{figschemlat}(c) are straightforward extensions of the lattice in Fig.~\ref{figschemlat}(a) by stacking and connecting chains of a similar structure as the chain in Fig.~\ref{figschemlat}(a). In Fig.~\ref{figschemlat}(b), we stack chains consisting of $A$ (red) and $B$ (green) sublattices, and connect them via $B'$ (blue) sublattices. On the resulting structure, we find that the wave function given by Eq.~(\ref{eqexactsolgen}) with $s = 0,1$ is an exact solution and corresponds to zero-dimensional corner modes or one-dimensional hinge states in which case each sublattice is associated with a one-dimensional periodic chain. While the lattice in Fig.~\ref{figschemlat}(b) corresponds to the Lieb lattice, in Sec.~\ref{secboundcodimtwo} we show an explicit example for the corner modes on the breathing kagome lattice, which only differs from the Lieb lattice by the inclusion of hopping terms between certain $B$ and $B'$ sublattices. In that section, we also solve explicitly for the hinge modes on the pyrochlore lattice. The Lieb lattice is treated explicitly in a forthcoming publication \cite{kunstvmiertbergholtz2}, where we solve the complete eigensystem by making use of a symmetry that relates the energies $E (k_x, k_y) = E (-k_x,k_y) = E (k_x,-k_y)$. To construct the lattice in Fig.~\ref{figschemlat}(c), we employ a similar construction as before, where we now stack chains consisting of $A$ and $B$, $A$ and $B'$, and $A$ and $B''$ (purple) sublattices in the $\vec{a}_{s+1}$ directions such that we can realize zero-dimensional corner states, which are given by Eq.~(\ref{eqexactsolgen}) with $s = 0,1,2$. In Sec.~\ref{secboundcodimthree}, we provide an explicit example for this schematic lattice in the form of the breathing pyrochlore model, which, similar to the breathing kagome lattice, is readily obtained by allowing for hoppings among certain $B$, $B'$, and $B''$ sublattices.

The lattice in Fig.~\ref{figschemlatmir} is also closely related to the chain in Fig.~\ref{figschemlat}(a), and consists of $A$, $B$, and $B'$ sublattices. As we do not allow for any hoppings between the $A$ sublattices, the wave function in Eq.~(\ref{eqexactsolgen}) with $s = 0, 1$ is an exact solution on this lattice. The chains formed by the $A$ and $B$ sublattices (in red and green) and the $B'$ sublattices (in blue) are connected in such a way that the lattice has a mirror symmetry around $m' = (M'+1)/2$, and consequently, $r'_i(\vec{k}_{||}) = -1, \, \forall i$. The modes captured by the exact wave-function solution in Eq.~(\ref{eqexactsolgen}) thus localize to the $A$ sublattice $\{m,m'\}= \{1,(M'+1)/2\}$ or $\{M,(M'+1)/2\}$ depending on whether $|r_i(\vec{k}_{||})|<1$ or $>1$, respectively. In Sec.~\ref{secmirrorsymm}, we show how we can use this construction to find exact solutions for hinge modes both in space and time.

In the following, we describe each of these lattices in great detail by studying their concomitant tight-binding Hamiltonians, writing down the relevant form of the exact wave-function solution in Eq.~(\ref{eqexactsolgen}), and deriving explicit expressions for $r_i^{(s)} (\vec{k}_{||})$.

\section{Exactly solvable boundary states with codimension one} \label{seccodimone}

\noindent In this section, we show how to find exact boundary-state wave functions on lattices with boundaries of codimension one as shown schematically in Fig.~\ref{figschemlat}(a). Solutions on lattices of this form are extensively described in Ref.~\onlinecite{kunsttrescherbergholtz} in the context of frustrated lattices, and are generalized here to a wider class of lattices. We supplement the general findings with an explicit example in the form of the SSH chain \cite{ssh}.

We start by describing generic, one-dimensional structures, which consist of alternating zero-dimensional $A$ and $B$ lattices, where the $A$ lattices are only coupled to adjacent $B$ lattices through the matrices $h_{A,B;\pm}$. Within each $A$ lattice, there are $n$ internal degrees of freedom, which may correspond to orbital, spin, or lattice-site degrees of freedom, and $h_A$ describes the physics within each $A$ lattice. We denote the corresponding creation operators with $c^\dagger_{A_i, m}$ with $m$ the unit-cell index and $i=1,\ldots ,n$. Each $B$ lattice hosts a single degree of freedom with energy $h_{B}$, and the corresponding creation operator is given by $b^{\dagger}_{m}$ with $m$ the unit cell index, which is the same as the unit-cell index of the $A$ lattice to its left. The Fourier-transformed creation operators are defined as 
\begin{equation*}
c^\dagger_{A_i,\vec{k}}=\frac{1}{\sqrt{L}}\sum_{m=1}^{L} e^{i m k_1}c^\dagger_{A_i, m},\quad
b^{\dagger}_{\vec{k}}=\frac{1}{\sqrt{L}}\sum_{m=1}^{L} e^{i m k_1}b^{\dagger}_{m},
\end{equation*}
where $L$ denotes the total number of unit cells in the periodic system, and $k_1 = \vec{k} \cdot \vec{a}_1$ with $\vec{a}_1$ the lattice vector shown in Fig.~\ref{figschemlat}(a). We find that the Bloch Hamiltonian is given by $H_{\vec{k}}=\Psi^\dagger_{\vec{k}}\mathcal{H}_{\vec{k}}\Psi_{\vec{k}} $ with $\Psi^\dagger_{\vec{k}} = (c^\dagger_{A_i,\vec{k}}, b^{\dagger}_{\vec{k}})$ and
\begin{equation*}
\mathcal{H}_{\vec{k}} = \begin{pmatrix}
h_A&h_{A,B;+}+e^{-ik_1}h_{A,B;-}\\
h_{A,B;+}^\dagger+e^{ik_1}h^\dagger_{A,B;-}&h_{B} 
\end{pmatrix}. \label{eqhamdodimone}
\end{equation*}
For simplicity, we have assumed that there is no hopping among the different $B$ sites, which does not restrict the generality of our conclusions as they are not altered by the inclusion of such hopping terms.
\begin{figure}[t]
{\includegraphics[width=\linewidth]{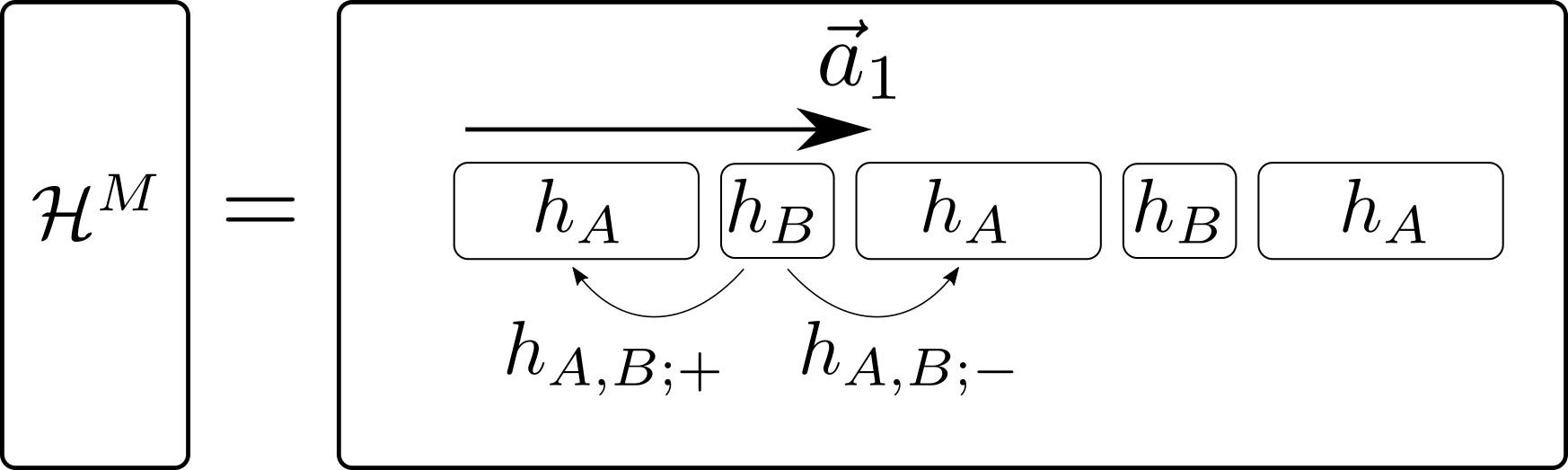}}
\caption{Schematic representation of the Hamiltonian $\mathcal{H}^M$ in Eq.~(\ref{eqgeneralhamonerow}).}
\label{fig1dschem}
\end{figure}

Next, we consider the corresponding system with open boundary conditions in $\vec{a}_1$ that starts and terminates with an $A$ lattice as shown in Fig.~\ref{figschemlat}(a). In this case, the Hamiltonian with a total of $M$ $A$ sublattices reads
\begin{equation*}
H^M=\Psi^\dagger \mathcal{H}^M\Psi,
\end{equation*}
where the row vector is given by $\Psi^\dagger=({\bf a}^\dagger_1,b^\dagger_1,{\bf a}^\dagger_2,\ldots,b^\dagger_{M-1},{\bf a}^\dagger_M)$ with ${\bf a}^\dagger_m=c^\dagger_{A_1},\ldots,c^\dagger_{A_n}$, and the matrix $\mathcal{H}^M$ is of the form
\begin{equation}
\mathcal{H}^M= \begin{pmatrix}
h_A & h_{A,B;+} & 0 & 0 & 0 \\
h_{A,B;+}^\dagger & h_B & h_{A,B;-}^\dagger  & 0 & 0 \\
0 & h_{A,B;-}  & h_A & \cdots & 0 \\
0 & 0 & \vdots & \ddots & h_{A,B;-}^\dagger \\
0 & 0 & 0 & h_{A,B;-}  & h_A
\end{pmatrix},  \label{eqgeneralhamonerow}
\end{equation}
which is shown schematically in Fig.~\ref{fig1dschem}. Given this structure, it is straightforward to construct exact wave-function solutions that localize at the boundaries formed by the $A$ sublattices at the ends. For simplicity, we assume that $h_A$ is diagonal, i.e., $h_A=\textrm{diag}(\epsilon_{A_1},\ldots,\epsilon_{A_n})$, which can be achieved by performing a unitary transformation. We continue by taking linear combinations of $c^\dagger_{A_i,m}|0\rangle$ and choose their weights in such a way that they interfere destructively on all the $B$ sublattices. Since the on-site energy is the same on all the $A$ lattices, i.e., $\epsilon_{A_i}$, we indeed find an eigenstate of the Hamiltonian. In particular, for each $i=1,\ldots,n$, we make the following ansatz, which corresponds to an exponentially decaying wave function
\begin{equation}
|\psi_{i}\rangle= \mathcal{N}_i \sum_{m=1}^{M} r^m_i c^\dagger_{A_i,m}|0\rangle, \label{eqeiggeneralstrucone}
\end{equation}
where $\mathcal{N}_i$ is the normalization constant
\begin{equation}
\mathcal{N}_i = \frac{1}{|r_{i}|} \sqrt{\frac{|r_i |^2-1}{(|r_i|^2)^{M}-1}}, \label{eqnormfact}
\end{equation} and $|\psi_{i}\rangle$ fulfils the energy eigenvalue equation
\begin{equation}
H^M|\psi_{i}\rangle=\epsilon_{A_i} |\psi_{i}\rangle .
\end{equation}
The explicit equation for $r_i$ follows from destructive interference on the $B$ lattices and is given by
\begin{equation}
(h^\dagger_{A,B;+})_{1,i}+r_i (h^\dagger_{A,B;-})_{1,i}=0, \qquad i=1,\ldots,n, \label{eqgeneralexprforr}
\end{equation}
where the labels $\{1,i\}$ refer to the matrix elements of the matrices $h^\dagger_{A,B;\pm}$ \footnote{In the rare event that $(h^\dagger_{A,B;-})_{1,i} = 0, \, \forall i$, we simply take $r_i=\infty$, which corresponds to a perfectly localized wave function on the end $m = M$.}. The function $r_i$ dictates the localization of the wave function, such that while the wave function is completely delocalized in the system when $|r_i| = 1$, we find that it localizes to the right end, $m = 1$, and left end, $m = M$, when $|r_i|<1$ and $|r_i|>1$, respectively. Note that we are able to solve this equation because the number of equations matches the number of variables $n$, which is why we restrict ourselves to the case where the $B$ lattices host a single orbital, i.e., $n_B = 1$. Moreover, the simple form of this equation for $r_i$ follows from writing the Hamiltonian $h_A$ in its diagonal basis. If $h_A$ is expressed in a different basis, we find that the ansatz in Eq.~(\ref{eqeiggeneralstrucone}) and the equation for $r_i$ in Eq.~(\ref{eqgeneralexprforr}) explicitly depend on the normalized eigenvectors $\phi_i$ of $h_A$ with eigenvalues $\epsilon_{A_i}$, such that they read
\begin{align}
&|\psi_i\rangle=\mathcal{N}_i \sum_{m=1}^{M} r^m_i \left(\sum_{j=1}^nc^\dagger_{A_j,m}\phi_{i,j}\right)|0\rangle, \\
&(h^\dagger_{A,B;+})\phi_i+r_i (h^\dagger_{A,B;-})\phi_i=0, \qquad i=1,\ldots,n,
\end{align}
where $\phi_{i,j}$ corresponds to the $j$th component of the $i$th eigenvector $\phi_i$.
 
 \begin{figure}[b]
{\includegraphics[width=\linewidth]{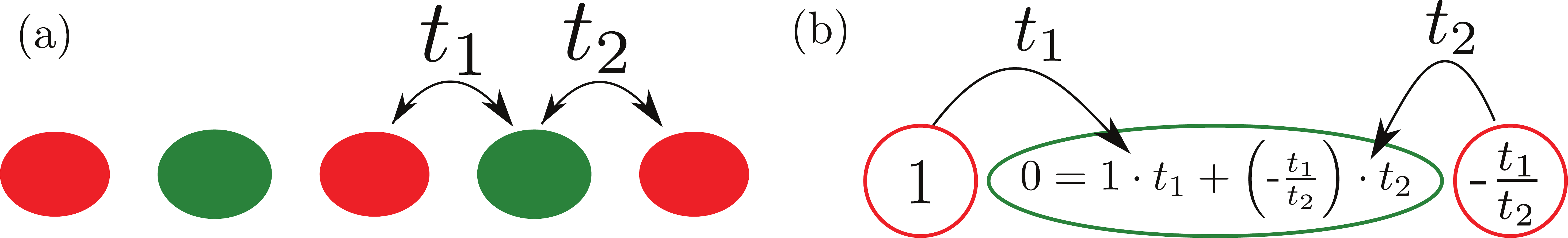}}
\caption{(a) Schematic depiction of the SSH chain with the $A$ and $B$ sublattices in red and green, respectively, and the nearest-neighbor hopping parameters $t_1$ and $t_2$. (b) Illustration of the wave function in Eq.~(\ref{eqeiggeneralstrucone}) with $r = -t_1/t_2$ for a chain with $M=2$ with the weight of the wave function on each site written inside the red ($A$) and green ($B$) circles to explicitly show the destructive interference on the $B$ (green) site.}
\label{figssh}
\end{figure}
\begin{figure}[t]
{\includegraphics[width=\linewidth]{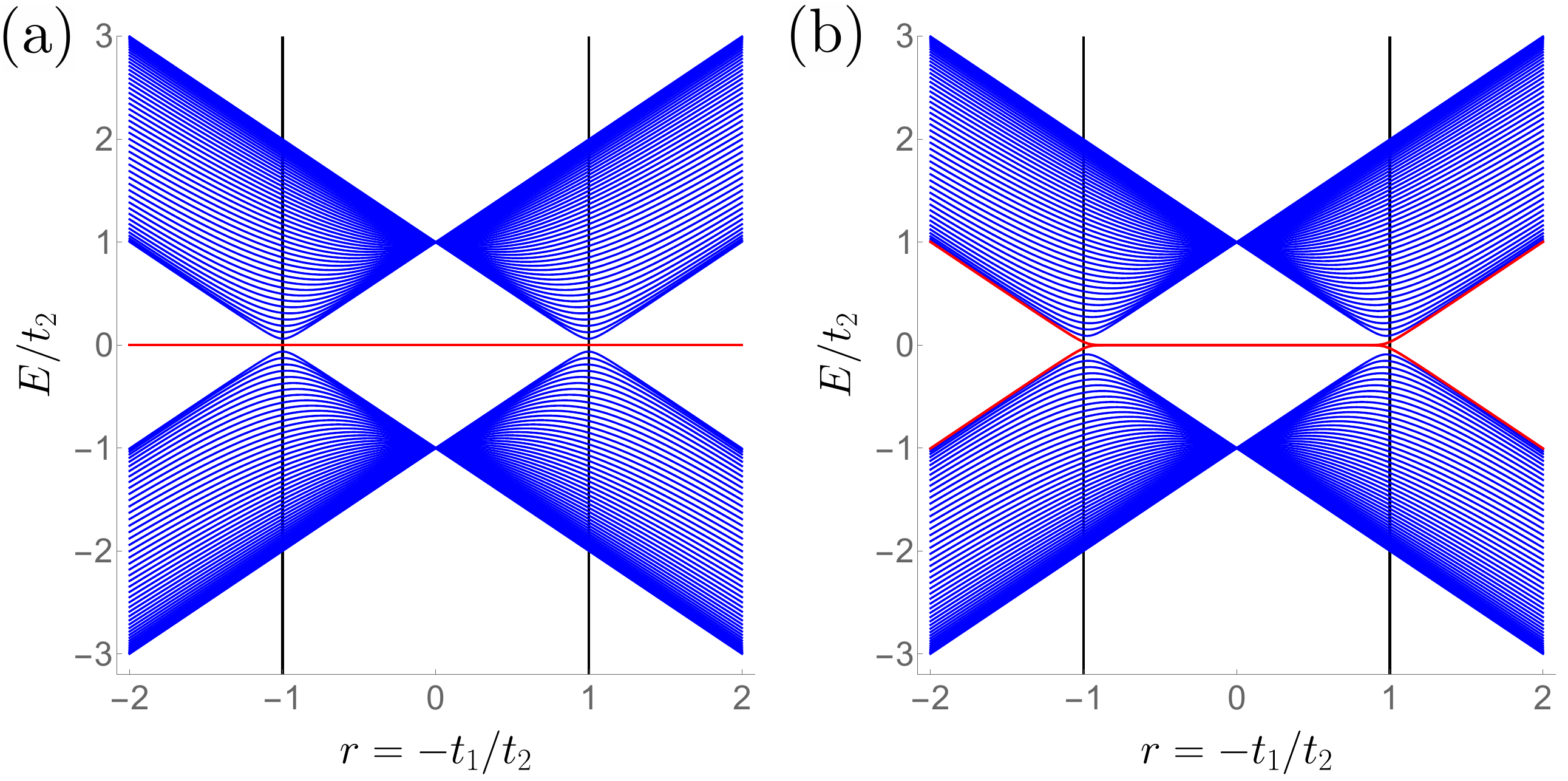}}
\caption{Band spectrum of the SSH model with $M = 50$ and (a) a broken unit cell (odd number of sites) and (b) an unbroken unit cell (even number of sites) with the bulk bands in blue and the end mode corresponding to the exact solution in Eq.~(\ref{eqeiggeneralstrucone}) in red. The black vertical lines indicate where $|r|=1$.}
\label{SSH_model}
\end{figure}

 
To elucidate these results, we work out an explicit example in the form of the SSH chain \cite{ssh}, which is also mentioned briefly in Ref.~\cite{kunstvmiertbergholtz}. This model consists of alternating $A$ and $B$ sites, which are coupled through the nearest-neighbor hopping parameters $t_1$ and $t_2$ as shown in Fig.~\ref{figssh}(a), such that the $A$ lattices host a single degree of freedom, i.e., $n=1$. The corresponding Bloch Hamiltonian reads
 \begin{equation*}
 \mathcal{H}_{\vec{k}}=-\begin{pmatrix}
 0&t_1+e^{-ik_1}t_2\\
 t_1+e^{ik_1}t_2&0
 \end{pmatrix}.
 \end{equation*}
For an open chain with $A$ sites at its ends, we find that the Hamiltonian is given by Eq.~(\ref{eqgeneralhamonerow}) with the exact wave-function solution in Eq.~(\ref{eqeiggeneralstrucone}). The terms in the Hamiltonian are
 \begin{equation*} 
 h_A = h_B=0, \qquad h_{A,B;+}=-t_1, \qquad h_{A,B;-}=-t_2,
 \end{equation*}
 where we use the notation introduced above such that Eq.~\eqref{eqgeneralexprforr} now reads
\begin{equation*}
t_1+r\, t_2=0,
\end{equation*}
where we have dropped the subscript $i=1$ of $r$ to simplify notation, and we find $r=-t_1/t_2$. The wave function for the open chain in Eq.~(\ref{eqeiggeneralstrucone}) with eigenvalue $h_A = 0$ is depicted in Fig.~\ref{figssh}(b), where the mechanism that results in the exact disappearance of the wave-function amplitude of the $B$ sites is explicitly illustrated. The band spectrum is shown in Fig.~\ref{SSH_model}(a), and we see that there is an exact zero-energy mode in red for all values of $r$. This result is in complete agreement with the well-known appearance of an exponentially localized end mode at $m = 1$ ($M$) for $|t_1| < |t_2|$ ($|t_1| > |t_2|$), and disappearance of this mode at $|t_1| = |t_2|$ ($|r|=1$) in the SSH chain with an odd number of sites (oddSSH).

We note that in the case with no broken unit cell, i.e., an even number of total sites (evenSSH), there are two zero-energy (up to algebraic finite-size corrections to the energy) end modes when $|t_1| < |t_2|$ and no end modes when $|t_1| \geq |t_2|$ as shown in Fig.~\ref{SSH_model}(b). In the case $|t_1| < |t_2|$, the two ends of the evenSSH chain can locally be mapped onto the end $m=1$ of the oddSSH chain, which indeed features an end mode when $|t_1| < |t_2|$. This mapping breaks down, however, when $|t_1| \geq |t_2|$ as the putative solutions interfere strongly with each other due to their increasing amplitude away from their respective ends. In this case, the exact solutions are no longer good approximate eigenstates of the even length system.
From the preceding discussion, we observe that while our solutions are only exact on lattices that terminate with $A$ sublattices on both ends, the quantity $r_i$ accurately predicts the presence of boundary modes also in the case where there are no broken unit cells, i.e., a termination with an $A$ sublattice on one end and a $B$ sublattice on the other end, in which case $|r_i|<1$ ($|r_i|>1$) signals the presence (absence) of boundary modes on both boundaries simultaneously.

We emphasize that the method described in this section is not restricted to one-dimensional models and can be straightforwardly generalized to higher dimensions as long as the boundaries of the model have codimension one. Indeed, by allowing the $A$ and $B$ sublattices in Fig.~\ref{figschemlat}(a) to be $(d=1)$ or $(d=2)$ dimensional, we form $(D=2)$- and $(D=3)$-dimensional lattices on which we can realize edge and surface modes, respectively. In these cases, one performs a partial Fourier transformation to reduce the model to a family of one-dimensional chains parametrized by $\vec{k}_{||}$, which is the momentum parallel to the boundary. This leads to the following, generalized exact solutions of the form of Eq.~(\ref{eqexactsolgen}),
\begin{equation*}
\ket{\psi_i (\vec{k}_{||})} = \mathcal{N}_i (\vec{k}_{||}) \sum_{m = 1}^{M} \left[r_i (\vec{k}_{||}) \right]^{m}  c_{A_i, \vec{k}_{||}, m}^\dagger \ket{0},
\end{equation*}
which differs from Eq.~(\ref{eqeiggeneralstrucone}) only by the explicit inclusion of momentum dependence. In Ref.~\onlinecite{kunsttrescherbergholtz}, this equation was retrieved for the chiral edge states of a Chern insulators and the Fermi arcs of a Weyl semimetal on frustrated lattices with $n = 2$ and $3$, and $n_{B}=1$ (see also Table~\ref{tabdiffmodels}).

\section{Exactly solvable boundary states with codimension two} \label{secboundcodimtwo}
\begin{figure}[b]
{\includegraphics[width=\linewidth]{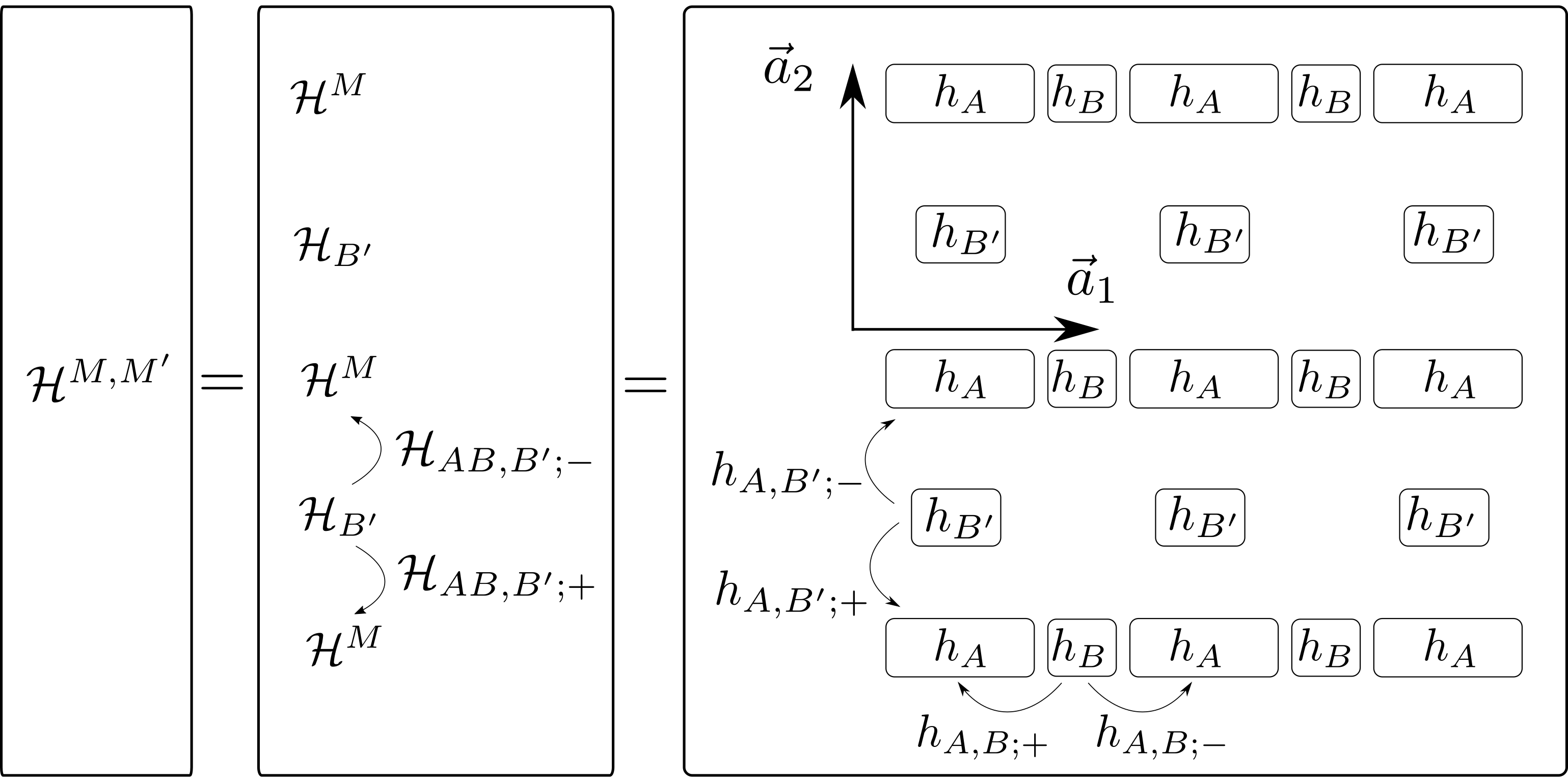}}
\caption{Schematic representation of the Hamiltonian $\mathcal{H}^{M, M'}$ in Eq.~(\ref{eqgeneralham}).}
\label{fig2dschem}
\end{figure}

\noindent We now shift our attention to models exhibiting solvable eigenstates that localize to boundaries with codimension two; these include the so-called second-order topological states. In particular, we find exact wave-function solutions for the zero-dimensional corner states in two-dimensional crystals of the type shown in Fig.~\ref{figschemlat}(b) and use this to show in detail how to obtain corner states in the breathing kagome lattice \cite{kunstvmiertbergholtz}. In addition to the examples mentioned in our previous paper \cite{kunstvmiertbergholtz}, we also include a new example of chiral, hinge states on the pyrochlore lattice.

In this discussion, we begin by considering two-dimensional models with zero-dimensional corners for the sake of simplicity but, as before, the method can be easily applied to higher dimensional systems with boundaries of codimension two of which an explicit example is shown at the end of this section. The idea is that we start with $M'$ one-dimensional chains of the form discussed in the previous section, which we then couple to each other in the
$\vec{a}_2$ direction by inserting intermediate $B'$ sublattices, such that the corners are formed by $A$ sublattices, as is shown in Fig.~\ref{figschemlat}(b). As before, the individual $A$ sublattices host $n$ degrees of freedom and are described by $h_A$, whereas both the $B$ and $B'$ sublattices host a single orbital with energies $h_B$ and $h_{B'}$, respectively. Moreover, the $A$ sublattices hybridize in the $\vec{a}_1$ ($\vec{a}_2$) direction only with the $B$ ($B'$) sublattices, which is described by the matrices $h_{A,B;\pm}$ ($h_{A,B';\pm}$). The creation operators on the $A$ sublattices are denoted by $c^\dagger_{A_i,m,m'}$ with $i=1,\ldots ,n$ and with $m$ and $m'$ the unit-cell indices. The creation operators on the $B$ ($B'$) sites read $b^\dagger_{m,m'}$ ($b'^\dagger_{m,m'}$), where each $B$ ($B'$) sublattice has the same unit-cell index as the $A$ sublattice to its left (below). The corresponding Fourier-transformed creation operators are given by
\begin{align}
&c^\dagger_{A_i,\vec{k}}=\frac{1}{\sqrt{LL'}}\sum_{m=1}^L\sum_{m'=1}^{L'}e^{i mk_1+i m'k_2}c^\dagger_{A_i,m,m'}, \label{eqfouriertransformc} \\
&b^\dagger_{\vec{k}}=\frac{1}{\sqrt{LL'}}\sum_{m=1}^L\sum_{m'=1}^{L'}e^{i mk_1+i m'k_2}b^\dagger_{m,m'}, \label{eqfouriertransformcone} \\
&b'^\dagger_{\vec{k}}=\frac{1}{\sqrt{LL'}}\sum_{m=1}^L\sum_{m'=1}^{L'}e^{i mk_1+i m'k_2}b'^\dagger_{m,m'}, \label{eqfouriertransformctwo}
\end{align}
where $L$ and $L'$ denote the total number of unit cells in the periodic system in the directions $\vec{a}_1$ and $\vec{a}_2$, respectively, and $k_i = \vec{k} \cdot \vec{a}_i$. Using this notation, we find that the full Bloch Hamiltonian is given by
$H_{\vec{k}}=\Psi^\dagger_{\vec{k}}\mathcal{H}_{\vec{k}}\Psi_{\vec{k}} $ with  $\Psi^\dagger_{\vec{k}} = (c^\dagger_{A_i,\vec{k}}, b^\dagger_{\vec{k}}, b'^\dagger_{\vec{k}})$ and
\begin{align}
\mathcal{H}_{\vec{k}} &= \begin{pmatrix}
h_A&h_{A,B}&h_{A,B'}\\
h_{A,B}^\dagger&h_B&0\\
h_{A,B'}^\dagger&0&h_{B'}
\end{pmatrix}, \label{eqblochhamcodimtwo} \\
h_{A,B} &\equiv h_{A,B;+}+e^{-ik_1}h_{A,B;-}, \label{eqblochhamcodimtwotwotwo} \\
h_{A,B'} &\equiv h_{A,B';+}+e^{-ik_2}h_{A,B';-}. \label{eqblochhamcodimtwotwo}
\end{align}
As in the previous section, we do not consider hoppings amongst the $B$ and $B'$ sublattices for reasons of simplicity while the inclusion of such terms does not alter any of our conclusions as is made explicit in the examples that follow below.

Next, we consider the same system with open boundary conditions in $\vec{a}_1$ and $\vec{a}_2$ with $A$ sublattices at its ends with $M$ ($M'$) $A$ blocks in the $\vec{a}_1$ ($\vec{a}_2$) direction as shown in Fig.~\ref{figschemlat}(b). For this model, the Hamiltonian reads
\begin{equation*}
H^{M,M'}=\Psi^\dagger \mathcal{H}^{M,M'}\Psi,
\end{equation*}
where the row vector $\Psi^\dagger$ is given by $
\Psi^\dagger=\left({\bf d}^\dagger_1, \,{\bf e}^\dagger_1, \, {\bf d}^\dagger_2, \,{\bf e}^\dagger_2, \, \ldots, \,{\bf e}^\dagger_{M'-1}, \, {\bf d}^\dagger_{M'}\right)$ with
\begin{align}
&{\bf d}^\dagger_{m'}= {\bf a}^\dagger_{1,m'}, \, b^\dagger_{1,m'}, \, {\bf a}^\dagger_{2,m'}, \, b^\dagger_{2,m'}, \, \ldots, \, b^\dagger_{M-1,m'}, {\bf a}^\dagger_{M,m'}, \label{eqdefvectors} \\ 
&{\bf a}^\dagger_{m,m'}=c^\dagger_{A_1,m,m'},\,\ldots,\,c^\dagger_{A_{n},m,m'}, \label{eqdefvectorstwo}\\ 
&{\bf e}^\dagger_{m'}=b'^\dagger_{1,m'}, \, \ldots, \,b'^\dagger_{M-1,m'}, \label{eqdefvectorsone}
\end{align}
and the matrix $\mathcal{H}^{M,M'}$ reads
\begin{align}
&\mathcal{H}^{M,M'} = \nonumber \\
&\begin{pmatrix}
\mathcal{H}^M & \mathcal{H}_{AB,B';+} & 0 & 0 & 0 \\
 \mathcal{H}_{AB,B';+}^{\dagger} & \mathcal{H}_{B'} &  \mathcal{H}_{AB,B';-} ^\dagger & 0 & 0 \\
0 & \mathcal{H}_{AB,B';-} &\mathcal{H}^M & \cdots & 0 \\
0 & 0 & \vdots & \ddots &\mathcal{H}_{AB,B';-} ^{\dagger} \\
0 & 0 & 0 & \mathcal{H}_{AB,B';-}  &\mathcal{H}^M
\end{pmatrix}, \label{eqgeneralham}
\end{align}
and is schematically shown in Fig.~\ref{fig2dschem}. The matrix $\mathcal{H}^M$, which is given by Eq.~\eqref{eqgeneralhamonerow} and shown schematically in Fig.~\ref{fig1dschem}, accounts for the physics within each row $m'$ composed of $M$ $A$ sublattices and $M-1$ $B$ sublattices. The matrix $\mathcal{H}_{B'}$ describes each row $m'$ composed of $M$ $B'$ sublattices, and can in principle assume any form. However, for simplicity we assume there are no interactions between $B'$ sublattices, such that its Hamiltonian is simply given by
\begin{equation*}
\mathcal{H}_{B'} = h_{B'} \, \mathbb{I}_{M \times M}. \label{eqhambprimeuseless}
\end{equation*}
The matrices $H_{AB,B';\pm}$ take care of the hybridization between neighboring rows, and read
\begin{align}
\mathcal{H}_{AB,B';+} &= \begin{pmatrix}
h_{A,B';+} & 0 & 0 & 0 \\
0 & 0 & 0 & 0 \\
0 & h_{A,B';+}  & 0 & 0 \\
0 & 0 & \cdots & 0 \\
0 & \vdots & \ddots & 0 \\
0 & 0 & 0 & h_{A,B';+} 
\end{pmatrix}, \label{eqperphamatobprimetwotwo} \\
\mathcal{H}_{AB,B';-} &= \begin{pmatrix}
h_{A,B';-} & 0 & 0 & 0 \\
0 & 0 & 0 & 0 \\
0 & h_{A,B';-} & \cdots & 0 \\
0 & \vdots & \ddots & 0 \\
0 & 0 & 0 & 0 \\
0 & 0 & 0 & h_{A,B';-}
\end{pmatrix}. \label{eqperphamatobprimetwo}
\end{align}

In a similar spirit, we now construct exact wave functions that localize to the corners, where we again assume that $h_A$ is written in a diagonal form, i.e., $h_A=\textrm{diag}(\epsilon_{A_1},\ldots,\epsilon_{A_n})$, such that the corner modes have energies $\epsilon_{A_i}$. From the previous section, we know how to construct the exact wave-function solutions for each single row $m'$, which localize to the left or right ends of the row, i.e., $m = 1$ or $M$, respectively. These $n$ wave-function solutions are given by
\begin{equation*}
|\Psi_{i;m'}\rangle= \mathcal{N}_i \sum_{m=1}^{M} r^m_i c^\dagger_{A_i,m,m'}|0\rangle, \label{eqeiggeneralstruconeextraindex}
\end{equation*}
which differ from the solutions in Eq.~(\ref{eqeiggeneralstrucone}) by the introduction of the unit-cell index $m'$ and where $\mathcal{N}_i$ and $r_i$ are given by Eqs.~\eqref{eqnormfact} and \eqref{eqgeneralexprforr}, respectively. Next, we go one step further by taking linear combinations of $|\Psi_{i;m'}\rangle$ and choosing their weights such that there is also destructive interference on the $B'$ sublattices. This leads to the following ansatz,
\begin{align}
|\Psi_{i}\rangle &= \mathcal{N}'_i \sum_{m'=1}^{M'} r'^{m'}_i |\Psi_{i;m'}\rangle \nonumber \\
&= \mathcal{N}_i \, \mathcal{N}'_i \sum_{m=1}^{M} \sum_{m'=1}^{M'} r^m_i r'^{m'}_i c^\dagger_{A_i,m,m'}|0\rangle, \label{eqmainresultinsupmat}
\end{align}
where the normalization constant $\mathcal{N}'_i$ is given by Eq.~\eqref{eqnormfact} with $r_i \rightarrow r'_i$ and $M \rightarrow M'$. The value of $r'_i$ is found by solving the following equation that corresponds to destructive interference on the $B'$ sites
\begin{equation}
(h^\dagger_{A,B';+})_{1,i}+r'_i (h^\dagger_{A,B';-})_{1,i}=0, \qquad  i=1,\ldots,n ,\label{eqgeneralexprforrprecise}
\end{equation}
such that $|\Psi_{i}\rangle$ indeed is a solution to the eigenvalue equation
\begin{equation}
H^{M,M'}|\Psi_{i}\rangle=\epsilon_{A_i} |\Psi_{i}\rangle .
\end{equation}
These results correspond to what was found in Ref.~\onlinecite{kunstvmiertbergholtz}, and we find that these wave functions remain exact as long as one does not include {\it direct hoppings} between the $A$ sublattices, while one is free to include any type of disorder or hoppings within each of the sublattices $B$ and $B'$.

 \begin{figure*}[t]
{\includegraphics[width=\linewidth]{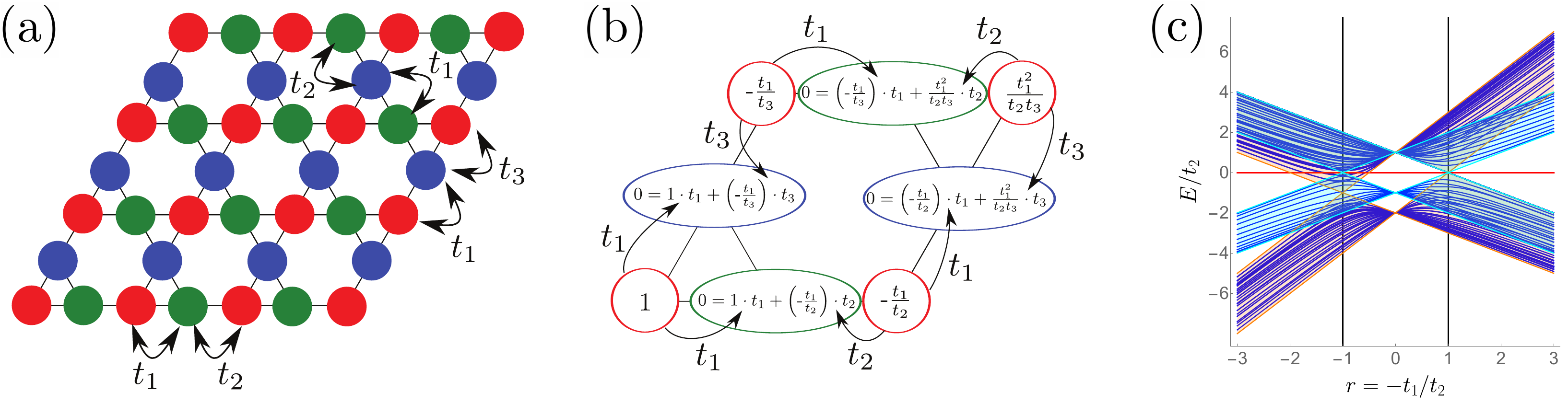}}
\caption{(a) Schematic depiction of the breathing kagome model with the $A$, $B$, and $B'$ sites in red, green, and blue, respectively, and the nearest-neighbor hopping parameters $t_1$, $t_2$, and $t_3$. (b) Illustration of the exact wave function in Eq.~(\ref{eqmainresultinsupmat}) with $r = -t_1/t_2$ and $r' = -t_1/t_3$ for a lattice with $M=M'=2$ with the weight of the wave function on each site written inside the red ($A$), green ($B$), and blue ($B'$) circles explicitly revealing the destructive interference on the $B$ (green) and $B'$ (blue) sites. (c) Energy spectrum with $M = M' = 7$ and $t_3/t_2 = 1$ such that $r = r' = -t_1/t_2$ with the bulk bands in blue and the corner mode corresponding to the exact solution in red. The bands in the orange and cyan shaded areas originate from the bulk and edge spectrum, respectively. The black vertical lines indicate where $|r|=1$.}
\label{figkagome}
\end{figure*} 
To illustrate these results, we consider the breathing kagome model shown in Fig.~\ref{figkagome}(a), which was also studied in Refs.~\onlinecite{kunstvmiertbergholtz, ezawapap, xuxuewan}. The arrangement of the sublattices $A$, $B$, and $B'$ shown in red, green, and blue, respectively, follows that of the lattice in Fig.~\ref{figschemlat}(b) and the only difference is the coupling between certain $B$ and $B'$ sublattices. The $A$ sublattices only couple to the neighboring $B$ ($B'$) sublattices with alternating, nearest-neighbor hoppings $t_1$ and $t_2$ ($t_1$ and $t_3$), while neighboring $B$ and $B'$ sites are coupled via alternating hoppings $t_1$ and $t_2$. Note that these hoppings between $B$ and $B'$ can be arbitrarily chosen as they do not influence the form of the exact wave-function solution. The corresponding Bloch Hamiltonian reads
\begin{align}
\mathcal{H}_{\vec{k}} &= \begin{pmatrix}
h_A&h_{A,B}&h_{A,B'}\\
h_{A,B}^\dagger&h_B&h_{B,B'}\\
h_{A,B'}^\dagger&h_{B,B'}^\dagger&h_{B'}
\end{pmatrix}, \label{eqblochhamcodimtwolarger} \\
h_{B,B'} &\equiv h_{B,B';+}+e^{-i (k_2 - k_1)}h_{B,B';-}, \label{eqblochhamcodimtwolargertwo}
\end{align}
which closely resembles the Hamiltonian in Eq.~(\ref{eqblochhamcodimtwo}), and $h_{A,B}$ and $h_{A,B'}$ given in Eqs.~(\ref{eqblochhamcodimtwotwotwo}) and (\ref{eqblochhamcodimtwotwo}), with
\begin{align}
& h_A = h_B = h_{B'} = 0, \qquad h_{A,B;-} = h_{B,B';-} =  - t_2, \nonumber \\
& h_{A,B;+} = h_{A,B';+} = h_{B,B';+} =  - t_1, \quad h_{A,B';-} = - t_3. \label{eqhamtermsbreathingkagome}
\end{align}
Considering open boundary conditions in the $\vec{a}_1$ and $\vec{a}_2$ direction, we find that the Hamiltonian is given by
Eq.~(\ref{eqgeneralham}) with $\mathcal{H}_{AB,B';\pm}$ reading
\begin{align}
\mathcal{H}_{AB,B';+} &= \begin{pmatrix}
h_{A,B';+} & 0 & 0 & 0 \\
h_{B,B';+} & 0 & 0 & 0 \\
0 & h_{A,B';+}  & 0 & 0 \\
0 & h_{B,B';+} & \cdots & 0 \\
0 & \vdots & \ddots & 0 \\
0 & 0 & 0 & h_{A,B';+} 
\end{pmatrix}, \label{eqperphamatobprimetwotwoextra} \\
\mathcal{H}_{AB,B';-} &= \begin{pmatrix}
h_{A,B';-} & 0 & 0 & 0 \\
0 & h_{B,B';-} & 0 & 0 \\
0 & h_{A,B';-} & \cdots & 0 \\
0 & \vdots & \ddots & 0 \\
0 & 0 & 0 & h_{B,B';-} \\
0 & 0 & 0 & h_{A,B';-}
\end{pmatrix}, \label{eqperphamatobprimetwoextra}
\end{align}
which differ from the matrices in Eqs.~(\ref{eqperphamatobprimetwotwo}) and (\ref{eqperphamatobprimetwo}) by the inclusion of extra hopping terms between certain $B$ and $B'$ sublattices. The exact wave function is given in Eq.~(\ref{eqmainresultinsupmat}) and is unaltered by the extra hopping terms as it only depends on the details of the Hamiltonian on sublattice $A$ and the hoppings between the $A$ and $B$ ($B'$) sites. The terms in the Hamiltonian are given in Eq.~(\ref{eqhamtermsbreathingkagome}), such that Eqs.~\eqref{eqgeneralexprforr} and \eqref{eqgeneralexprforrprecise}, which correspond to destructive interference on the $B$ and $B'$ sites, respectively, read
\begin{equation}
t_1+r \, t_2=0, \qquad \qquad t_1+r' \, t_3=0,
\end{equation}
where we have dropped the subscript $1$ on $r$ and $r'$ to simplify notation, and which yields $r=-t_1/t_2$ and $r'=-t_1/t_3$. The exact wave function is shown explicitly in Fig.~\ref{figkagome}(b) and the energy spectrum for $t_2 = t_3$ is shown in Fig.~\ref{figkagome}(c). The eigenenergy of the exact solution is $h_A = 0$ as shown in red in Fig.~\ref{figkagome}(c). Depending on the values of $|r|$ and $|r'|$, the zero-energy mode localizes to one of the four corners of the kagome lattice, i.e., the mode is localized to the corner $\{m,m'\} = \{1,1\}$, $\{M,1\}$, $\{1,M'\}$, or $\{M,M'\}$ when $\{|r|, |r'|\} = \{<1, <1\}$, $\{>1, <1\}$, $\{<1, >1\}$, or $\{>1, >1\}$, respectively, such that the zero-energy mode in the band spectrum in Fig.~\ref{figkagome}(c), which is plotted for $t_3/t_2=1$ such that $r = r'$, is localized to the corner $\{m,m'\} = \{1,1\}$ ($\{M,M'\}$) when $|r|<1$ ($>1$). Note that these results were also summarized in Ref.~\onlinecite{kunstvmiertbergholtz}. Indeed, the existence of the these corner modes has recently been confirmed in experiments \cite{xueyanggaochongzhang, niweineraluetal, elhassankunstmoritzandlerbergholtzbourenanne}.

\begin{figure*}[t]
{\includegraphics[width=1\linewidth]{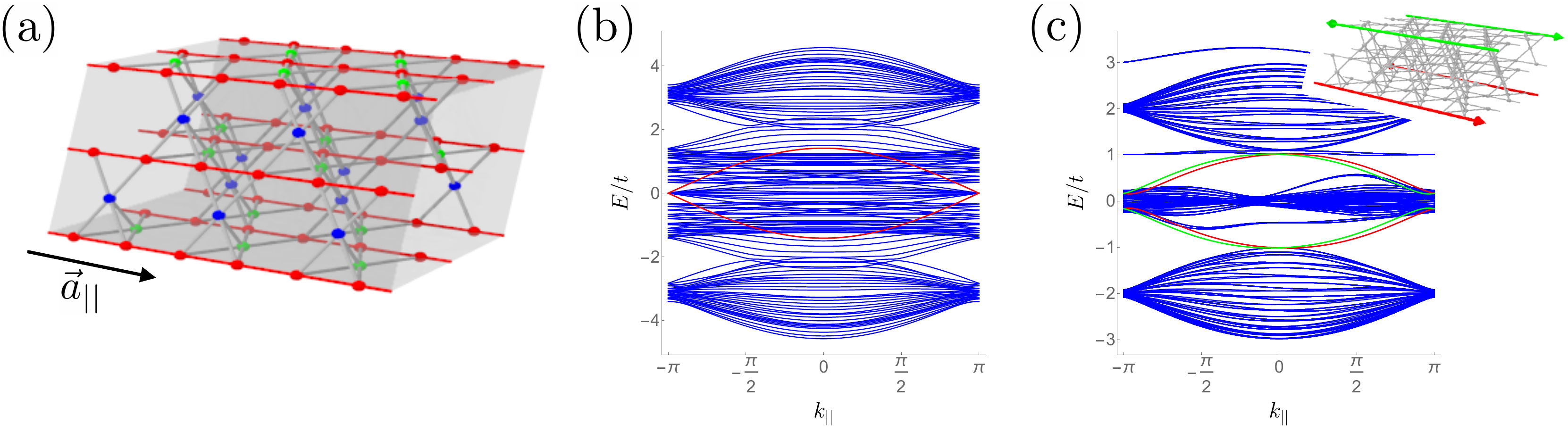}}
\caption{(a) Pyrochlore lattice with hinges parametrized by the lattice vector $\vec{a}_{||}$ and the $A$, $B$, and $B'$ sublattices shown in red, green, and blue, respectively. [(b), (c)] Energy spectrum for the model with an (b) all-in-all-out spin configuration on each tetrahedron with $t_\perp/t = t'_{\perp}/t = 1$ and $M = M' = 6$, and (c) a Kane-Mele model on each kagome layer with $t_\perp/t = (1/10) \, t'_{\perp}/t = t''_{\perp}/t =(1/6) V/t = (5/24) V_1 / t= 1$ and $M = M' = 10$, with the bulk bands in blue and the exact, hinge-state solution in red [and green in panel (c)]. The inset in panel (c) depicts the localization of the modes on the hinges.}
\label{fig_hinges_on_pyr_lattice_energy_bands}
\end{figure*}
Next, we consider the pyrochlore lattice shown in Fig.~\ref{fig_hinges_on_pyr_lattice_energy_bands}(a). This lattice can be interpreted as a dimensional extension of the kagome lattice in Fig.~\ref{figkagome}(a) by considering the $A$, $B$, and $B'$ sites in the latter as one-dimensional, periodic chains, which are connected to each other in such a way that they form the pyrochlore lattice. The general structure in Fig.~\ref{figschemlat}(b) thus again applies. We implement two different tight-binding models on this lattice of which the first realizes an all-in-all-out spin configuration on the tetrahedra, which is relevant for the pyrochlore irridates Eu$_2$Ir$_2$O$_7$ \cite{fujitakozukauchidatsukazakiarimakawasaki} and Nd$_2$Ir$_2$O$_7$ \cite{gallagheressermorrowdunsigerwilliamswoodwardmccombyang}. The Bloch Hamiltonian is given by Eq.~(\ref{eqblochhamcodimtwolarger}) supplemented by Eqs.~(\ref{eqblochhamcodimtwotwotwo}), (\ref{eqblochhamcodimtwotwo}), and (\ref{eqblochhamcodimtwolargertwo}) with
\begin{align}
&h_A =\frac{t}{\sqrt{2}}\begin{pmatrix}
0& 1 + e^{-i k_{||}}\\
1 +e^{i k_{||}} &0
\end{pmatrix}, \qquad h_B = h_{B'} = 0, \nonumber \\
&h_{A,B;+}=t_\perp\begin{pmatrix}(1+i)\\
(1-i)
\end{pmatrix}, \quad h_{A,B;-}=t_\perp\begin{pmatrix}(1+i)\\
(1-i)e^{i k_{||}}
\end{pmatrix}, \nonumber\\
&h_{A,B';+}=\frac{t'_{\perp}}{2}\begin{pmatrix}(1-i)\\
(1+i)
\end{pmatrix}, \quad h_{A,B';-}=\frac{t'_{\perp}}{2}\begin{pmatrix}(1-i)\\
(1+i)e^{i k_{||}}
\end{pmatrix},\nonumber \\
&h_{B,B';+} = h_{B,B';-} = \frac{t'_{\perp}}{\sqrt{2}}, \label{eqhamhingepyrone}
\end{align}
where $t$, $t_\perp$, and $t'_{\perp}$ are nearest-neighbor hopping parameters, and the model is periodic in $k_{||} = \vec{k} \cdot \vec{a}_{||}$ with $\vec{a}_{||}$ the lattice vector shown in Fig.~\ref{fig_hinges_on_pyr_lattice_energy_bands}(a). This Hamiltonian renders a Chern-insulating phase on each kagome-layer cut in the pyrochlore lattice, such that we may make use of the chiral edge states on the edges of each of these cuts to create a second-order topological insulators with hinge states. Assuming open boundary conditions in the $\vec{a}_1$ and $\vec{a}_2$ directions such that we obtain the lattice in Fig.~\ref{fig_hinges_on_pyr_lattice_energy_bands}(a), we obtain the Hamiltonian in Eqs.~(\ref{eqgeneralham}), (\ref{eqperphamatobprimetwotwoextra}), and (\ref{eqperphamatobprimetwoextra}) parametrized by $k_{||}$ with the Hamiltonian terms given in Eq.~(\ref{eqhamhingepyrone}). Note that we have not written $h_A$ in its diagonal form, such that we cannot simply use Eqs.~(\ref{eqgeneralexprforr}) and (\ref{eqgeneralexprforrprecise}) to find $r_i$ and $r'_i$, with $i=1,2$. Instead, we first find the eigenvectors of $h_A$, which are given by $\phi_{1}(k_{||})=(1,e^{i k_{||}/2})^T/\sqrt{2}$ and $\phi_{2}(k_{||})=(1,-e^{i k_{||}/2})^T/\sqrt{2}$ with eigenvalues $t\sqrt{2}\cos(k_{||}/2)$ and $-t\sqrt{2}\cos(k_{||}/2)$, respectively. In terms of these eigenvectors, we find that the equations corresponding to destructive interference on the $B$ and $B'$ sites read
\begin{equation*}
r_i(k_{||})=- \frac{h^\dagger_{A,B;+} \phi_i(k_{||})}{ h^\dagger_{A,B;-} \phi_i(k_{||})} 
, \qquad r'_i(k_{||})=- \frac{h^\dagger_{A,B';+} \phi_i(k_{||})}{ h^\dagger_{A,B';-} \phi_i(k_{||})},
\end{equation*}
with $i=1,2$, such that
\begin{align*}
r_\pm(k_{||})&= - \frac{(1 - i) \pm (1 + i) e^{i k_{||}/2}}{(1 - i) \pm (1 + i) e^{-i k_{||}/2}}, \\
 r'_\pm(k_{||})&=- \frac{(1 + i) \pm (1 - i) e^{i k_{||}/2}}{(1 + i) \pm (1 - i) e^{-i k_{||}/2}},
\end{align*}
where we have substituted the index $i$ for $\pm$ where $+$ ($-$) corresponds to $i = 1$ ($i = 2$). The energy spectrum is shown in Fig.~\ref{fig_hinges_on_pyr_lattice_energy_bands}(b) with the bands highlighted in red corresponding to the exact solution given in Eq.~(\ref{eqmainresultinsupmat}) with $r_i$ and $r'_i$ specified above. However, the bands are completely blurred by the bulk bands in blue, and we need to implement a more intricate model in order to isolate the chiral hinge states.

To this end, we continue by implementing a Kane-Mele-type Hamiltonian on each kagome layer \cite{kaneandmele, kaneandmeletwo}, such that each of the two sites in the $A$ lattices and the $B$ sites host two degrees of freedom instead of one, while the $B'$ sites still host a single orbital per site. The corresponding Bloch Hamiltonian is given by Eq.~(\ref{eqblochhamcodimtwolarger}) supplemented by Eqs.~(\ref{eqblochhamcodimtwotwotwo}), (\ref{eqblochhamcodimtwotwo}), and (\ref{eqblochhamcodimtwolargertwo}) with
\begin{align}
&h_A =\begin{pmatrix}
h_{A,+}&0 \\
0 & h_{A,-}
\end{pmatrix}, \nonumber \\
& h_{A,+} \equiv \begin{pmatrix}
V+t_2\sin(k_{||})&-i\frac{t}{2}(1+e^{-ik_{||}})\\
i\frac{t}{2}(1+e^{ik_{||}})&-V-t_2\sin(k_{||})
\end{pmatrix}, \\
& h_{A,-} \equiv \begin{pmatrix}
V-t_2\sin(k_{||})&i\frac{t}{2}(1+e^{-ik_{||}})\\
-i\frac{t}{2}(1+e^{ik_{||}})&-V+t_2\sin(k_{||})
\end{pmatrix}, \nonumber \\
&h_{A,B;+} =t_\perp\begin{pmatrix}
1 & 0\\
1 & 0 \\
0 & 1 \\
0 & 1
\end{pmatrix}, \quad h_{A,B;-}=t_\perp\begin{pmatrix}
1 & 0\\
e^{ik_{||}} & 0 \\
0 & 1 \\
0 & e^{ik_{||}}
\end{pmatrix}, \nonumber \\
& h_{A,B';+}= t'_{\perp}\begin{pmatrix}
0 \\
0 \\
1\\
1
\end{pmatrix}, \qquad
h_{A,B';-}=t'_{\perp}\begin{pmatrix}
1\\
e^{ik_{||}} \\
0 \\
0
\end{pmatrix},\nonumber
\end{align} \begin{align}
&h_{B,B';+} = \begin{pmatrix}
0 \\
t'_{\perp}
\end{pmatrix}, \qquad h_{B,B';-} = \begin{pmatrix}
t'_{\perp} \\
0
\end{pmatrix}, \nonumber\\
& h_B =\begin{pmatrix}
V_1&0\\
0&V_1
\end{pmatrix}, \qquad h_{B'} =
V_1, \label{blochhampyrwithkmmod}
\end{align}
where $t$, $t_\perp$ and $t'_{\perp}$ are nearest-neighbor hopping parameters, $t_2$ is a next-nearest-neighbor hopping parameter, and $V$ and $V_1$ are on-site potentials. Assuming open boundary conditions in $\vec{a}_1$ and $\vec{a}_2$ as before, we find the Hamiltonian given in Eqs.~(\ref{eqgeneralham})-(\ref{eqperphamatobprimetwo}) parametrized by $k_{||}$ with the Hamiltonian terms given in Eq.~(\ref{blochhampyrwithkmmod}), where the two degrees of freedom on the $B$ lattices turn the (creation) annihilation operator (${\bf b}^\dagger_{m,m'}$) ${\bf b}_{m,m'}$ into a (row) column vector of (creation) annihilation operators ($b^\dagger_{B_{i}, m,m'}$) $b_{B_{i},m,m'}$. Note that the Hamiltonian on each layer $m'$ simply realizes two Chern insulators with opposite Chern numbers that are decoupled from each other such that the exact wave-function solution in Eq.~(\ref{eqmainresultinsupmat}) still solves our model. The eigenvectors of $h_A$ are given by $\phi_{i}(k_{||})$ with $i=1,2,3,4$, where each eigenvector only has two nonzero components out of four with $i = 1,2$ ($3,4$) having a non-zero component in the first and second (third and fourth) rows, such that
\begin{align*}
r_i(k_{||})&=- \frac{(h^\dagger_{A,B;+})_{j,i} \, \phi_i(k_{||})}{(h^\dagger_{A,B;-})_{j,i} \, \phi_i(k_{||})} 
, \\
r'_i(k_{||})&=- \frac{(h^\dagger_{A,B';+})_{1,i} \, \phi_i(k_{||})}{ (h^\dagger_{A,B';-})_{1,i} \, \phi_i(k_{||})},
\end{align*}
where $ j =1$ ($j=2$) if $i = 1,2$ ($i=3,4$). From the explicit form of $\phi_{i}(k_{||})$ and $h^\dagger_{A,B';\pm}$, we find that $r'_i (k_{||}) = 0$ ($= -\infty$) if $i = 1,2$ ($i = 3,4$), such that the wave-function solution in Eq.~(\ref{eqmainresultinsupmat}) reduces to
\begin{align}
&\ket{\Psi_{i} (k_{||})} = \nonumber \\
& \quad \begin{cases}
               \mathcal{N}_{i} (k_{||})  \sum_{m=1}^{M} (r_{i}(k_{||}))^m c^\dagger_{A_i,m,m'=1}\ket{0}, \quad i = 1,2, \\
               \mathcal{N}_{i} (k_{||})  \sum_{m=1}^{M} (r_{i}(k_{||}))^m c^\dagger_{A_i,m,m'=M'}\ket{0}, \quad i = 3,4,
            \end{cases} \label{eqsolhingemodeskmpyr}
\end{align}
and we find that two of the exact solutions are precisely localized to the layer $m'=1$ while the other two solutions are localized to the layer $m' = M'$. The band spectrum for this model is shown in Fig.~\ref{fig_hinges_on_pyr_lattice_energy_bands}(c), where the perpendicular hopping between the layers $m'$ is taken to be $t''_{\perp}$ on the surfaces $\{m,m'\}=\{\tilde{m},1\}$, and $\{\tilde{m},M'\}$ with $\tilde{m} \in [1, M]$, such that the chiral states on the edges of each individual kagome layer formed by $A$ and $B$ sublattices are hybridized away and only four chiral states remain inside the gap whose bands are shown in red and green in Fig.~\ref{fig_hinges_on_pyr_lattice_energy_bands}(c) and are well separated from the bulk. By accessing the bands in the gap or by making use the solutions in Eq.~(\ref{eqsolhingemodeskmpyr}) explicitly, we can show that each of the chiral modes lives on a different hinge resulting in the localization shown in the inset in Fig.~\ref{fig_hinges_on_pyr_lattice_energy_bands}(c).

\section{Exactly solvable boundary states with codimension three} \label{secboundcodimthree}
\noindent Finally, we repeat the same steps as before to find exact solutions for the zero-dimensional corner modes of the three-dimensional model shown in Fig.~\ref{figschemlat}(c), which is a dimensional extension of the general lattice in Fig.~\ref{figschemlat}(b). We demonstrate the validity of our findings by applying it to retrieve zero-energy corner modes in the breathing pyrochlore lattice.

\begin{figure*}[t]
  \adjustbox{trim={0\width} {0\height} {0\width} {0\height},clip}
{\includegraphics[width=\linewidth]{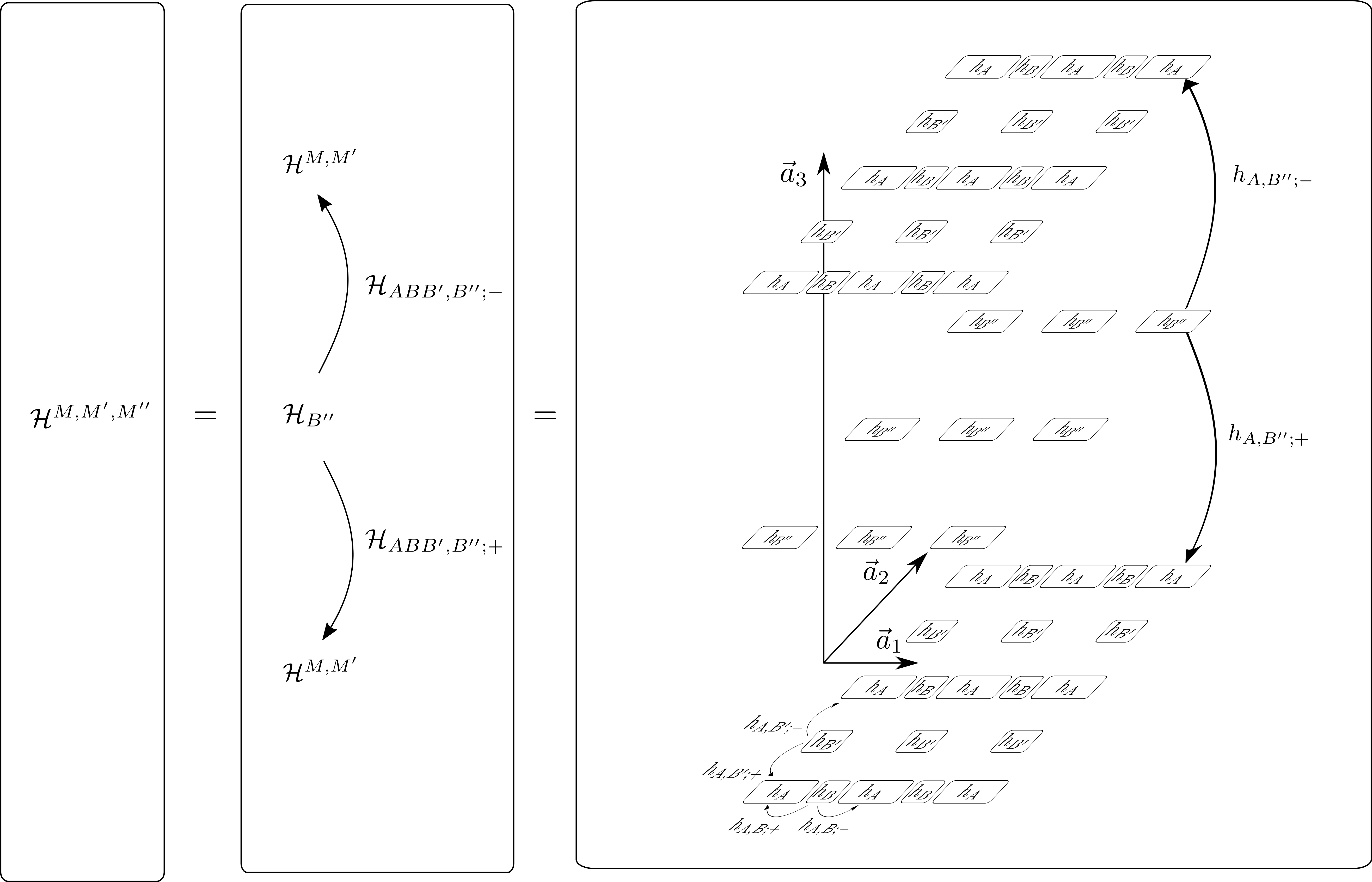}}
\caption{Schematic representation of the Hamiltonian $\mathcal{H}^{M,M',M''}$ in Eq.~(\ref{eqgeneralham}). }
\label{fig3dschem}
\end{figure*}

In a similar spirit as before, we start with $M''$ copies of the two-dimensional crystal discussed in the previous section for which we can find exact boundary states. We couple these models by inserting intermediate $B''$ sublattices, which host a single orbital only at an energy $h_{B''}$ as shown in Fig.~\ref{figschemlat}(c). Again, it is crucial that the $A$ sublattices only talk to neighboring $B$, $B'$, and $B''$ sublattices. The creation operators on the $A$ sublattice are denoted by $c^\dagger_{A_i,m,m',m''}$, where $i=1,\ldots,n$ labels the internal degrees of freedom, and $m,m',m''$ specifies the unit cell. In a similar fashion, we denote the creation operators corresponding to the $B$, $B'$, and $B''$ sublattices with $b^\dagger_{m,m',m''}$, $b'^\dagger_{m,m',m''}$, and $b''^\dagger_{m,m',m''}$, respectively, where the $B$, $B'$, and $B''$ sublattices have the same unit-cell index as the $A$ sublattice to their left in $\vec{a}_1$, $\vec{a}_2$ and $\vec{a}_3$, respectively, with $a_i$ the lattice vectors shown in Fig.~\ref{fig3dschem}. The corresponding Fourier-transformed creation operators are given by
\begin{align*}
&c^\dagger_{A_i,\vec{k}}= \\
&\frac{1}{\sqrt{LL'L''}}\sum_{m=1}^L\sum_{m'=1}^{L'}\sum_{m''=1}^{L''}e^{i mk_1+i m'k_2+i m''k_3}c^\dagger_{A_i,m,m',m''},
\end{align*} \begin{align*}
& b^\dagger_{\vec{k}}= \\
 &\frac{1}{\sqrt{LL'L''}}\sum_{m=1}^L\sum_{m'=1}^{L'}\sum_{m''=1}^{L''}e^{i mk_1+i m'k_2+i m''k_3}b^\dagger_{m,m',m''},
 \end{align*} \begin{align*}
&b'^\dagger_{\vec{k}}= \\
&\frac{1}{\sqrt{LL'L''}}\sum_{m=1}^L\sum_{m'=1}^{L'}\sum_{m''=1}^{L''}e^{i mk_1+i m'k_2+i m''k_3}b'^\dagger_{m,m',m''},
\end{align*} \begin{align*}
&b''^\dagger_{\vec{k}}= \\
&\frac{1}{\sqrt{LL'L''}}\sum_{m=1}^L\sum_{m'=1}^{L'}\sum_{m''=1}^{L''}e^{i mk_1+i m'k_2+i m''k_3}b''^\dagger_{m,m',m''},
\end{align*}
where $L$, $L'$, and $L''$ denote the total number of unit cells in the direction $\vec{a}_1$, $\vec{a}_2$, and $\vec{a}_3$, respectively, and $k_i = \vec{k} \cdot \vec{a}_i$. This leads to the Bloch Hamiltonian $H_{\vec{k}}=\Psi^\dagger_{\vec{k}}\mathcal{H}_{\vec{k}}\Psi_{\vec{k}} $ with  $\Psi^\dagger_{\vec{k}} = (c^\dagger_{A_i,\vec{k}}, b^\dagger_{\vec{k}}, b'^\dagger_{\vec{k}}, b''^\dagger_{\vec{k}})$ and
\begin{align}
\mathcal{H}_{\vec{k}} &= \begin{pmatrix}
h_A&h_{A,B}&h_{A,B'}&h_{A,B''}\\
h_{A,B}^\dagger&h_B&0&0\\
h_{A,B'}^\dagger&0&h_{B'}&0 \\
h^\dagger_{A,B''} & 0 & 0 & h_{B''}
\end{pmatrix}, \label{eqblochhamcodimthree} \\
h_{A,B''} &\equiv h_{A,B'';+}+e^{-i k_3}h_{A,B'';-}, \label{eqblochhamcodimthreetwo}
\end{align}
with $h_{A,B}$ and $h_{A,B'}$ given in Eqs.~(\ref{eqblochhamcodimtwotwotwo}) and (\ref{eqblochhamcodimtwotwo}), respectively.

Next, we provide the Hamiltonian that describes the same system with open boundary conditions in all three directions as shown in Fig.~\ref{figschemlat}(c).
The Hamiltonian is given by $H^{M\times M' \times M''} = \Psi^\dagger \mathcal{H}^{M,M',M''} \Psi$ with a total of $(M\times M' \times M'')$ $A$ lattices, where the row vector reads $
\Psi^\dagger=\left({\bf c}^\dagger_1, \,{\bf g}^\dagger_1, \, {\bf c}^\dagger_2, \,{\bf g}^\dagger_2, \, \ldots, \,{\bf g}^\dagger_{M''-1}, \, {\bf c}^\dagger_{M''}\right)$ with
\begin{align*}
&{\bf c}^\dagger_{m''}= {\bf d}^\dagger_{1,m''}, \,{\bf e}^\dagger_{1,m''}, \, {\bf d}^\dagger_{2,m''}, \,{\bf e}^\dagger_{2,m''}, \, \ldots, \,{\bf e}^\dagger_{M'-1,m''}, \, {\bf d}^\dagger_{M',m''}, \\ 
&{\bf g}^\dagger_{m''}={\bf h}^\dagger_{1,m''}, \, \ldots, \,{\bf h}^\dagger_{M-1,m''}, \\
&{\bf h}^\dagger_{m',m''}=b''^\dagger_{1,m',m''}, \, \ldots, \,b''^\dagger_{M-1,m',m''},
\end{align*}
with ${\bf d}^\dagger_{m', m''}$ and ${\bf e}^\dagger_{m', m''}$ defined in Eqs.~(\ref{eqdefvectors}) and (\ref{eqdefvectorsone}), respectively, with an extra label $m''$ and the matrix $\mathcal{H}^{M,M',M''}$ is given by
\begin{widetext}
\begin{equation}
\mathcal{H}^{M,M',M''} = \begin{pmatrix}
\mathcal{H}^{M,M'} & \mathcal{H}_{ABB',B'';+} & 0 & 0 & 0 \\
\mathcal{H}_{ABB',B'';+}^{\dagger} & \mathcal{H}_{B''} & \mathcal{H}_{ABB',B'';-}^{\dagger} & 0 & 0 \\
0 & \mathcal{H}_{ABB',B'';-} & \mathcal{H}^{M,M'} & \cdots & 0 \\
0 & 0 & \vdots & \ddots & \mathcal{H}_{ABB',B'';-}^{\dagger} \\
0 & 0 & 0 & \mathcal{H}_{ABB',B'';-} & \mathcal{H}^{M,M'}
\end{pmatrix}, \label{eqgeneralhampyrcorner}
\end{equation}
\end{widetext}
and schematically shown in Fig.~\ref{fig3dschem}. The matrices  $\mathcal{H}^{M,M'}$ and $\mathcal{H}_{B''} $ on the diagonal correspond to the physics within each two-dimensional lattice labeled by $m''$ that are stacked in the $\vec{a}_3$ direction, and are given by Eq.~(\ref{eqgeneralham}) and $\mathcal{H}_{B''} = \mathbb{I}_{M M' \times M M'} \otimes h_{B''}$, where we assumed no hopping between $B''$ lattices. The matrices $ \mathcal{H}_{ABB',B'';\pm}$ hybridize neighboring planes and read
\begin{equation}
\mathcal{H}_{ABB',B'';\pm} =  \begin{pmatrix}
h_{AB,B'';\pm} & 0 & 0 & 0 \\
0& 0 & 0 & 0 \\
0 & h_{AB,B'';\pm} & 0 & 0 \\
0 & 0& \cdots & 0 \\
0 & \vdots & \ddots & 0\\
0 & 0 & 0 & h_{AB,B'';\pm}
\end{pmatrix}, \label{eqhoppingbetweenbbpprime}
\end{equation}
with
\begin{equation}
h_{AB,B'';\pm} = \begin{pmatrix}
h_{A,B'';\pm} & 0 & 0 & 0 \\
0& 0 & 0 & 0 \\
0 & h_{A,B'';\pm} & 0 & 0 \\
0 & 0& \cdots & 0 \\
0 & \vdots & \ddots & 0 \\
0 & 0 & 0 & h_{A,B'';\pm}
\end{pmatrix}. \label{eqhoppingbetweenbbpprimetwo}
\end{equation}
Again, we have not included hoppings that hybridize $B$, $B'$, and $B''$ sublattices for the sake of simplicity.

We continue by finding exact corner-state solutions for this structure. Note that in the previous section we found exact wave-function solutions within each single plane $m''$, which typically localize to one of the four corners. For the three-dimensional system, these are given by
\begin{equation*}
|\Psi_{i;m''}\rangle = \mathcal{N}_i \, \mathcal{N}'_i \sum_{m=1}^{M} \sum_{m'=1}^{M'} r^m_i r'^{m'}_i c^\dagger_{A_i,m,m',m''}|0\rangle,
\end{equation*}
with $i=1,\ldots,n$, and differ only from Eq.~(\ref{eqmainresultinsupmat}) by an extra unit-cell index $m''$. As before, we take linear combinations of these wave functions such that they also interfere destructively on the $B''$ sublattices. This leads to the following wave-function solutions
\begin{align}
&|\Psi_i\rangle= \mathcal{N}''_i \sum_{m''=1}^{M''}r''^{m''}_i|\Psi_{i,m''}\rangle \nonumber \\
&= \mathcal{N}_i \, \mathcal{N}'_i \, \mathcal{N}''_i \sum_{m=1}^{M}\sum_{m'=1}^{M'}\sum_{m''=1}^{M''}r^{m}_ir'^{m'}_ir''^{m''}_ic^\dagger_{A_i,m,m',m''}|0\rangle, \label{eqcornerstatethreedim}
\end{align}
where $ \mathcal{N}''_i $ is the normalization constant given by Eq.~\eqref{eqnormfact} with $r_i \rightarrow r''_i$ and $M \rightarrow M''$, and $r''_i$ solves the following equation:
\begin{equation}
(h^\dagger_{A,B'';+})_{1,i}+r''_i (h^\dagger_{A,B'';-})_{1,i}=0, \qquad  i=1,\ldots,n, \label{eqletslabelanotherone}
\end{equation}
ensuring
\begin{equation}
H^{M,M',M''}|\Psi_{i}\rangle=\epsilon_{A_i} |\Psi_{i}\rangle .
\end{equation}

\begin{figure*}[t]
{\includegraphics[width=\linewidth]{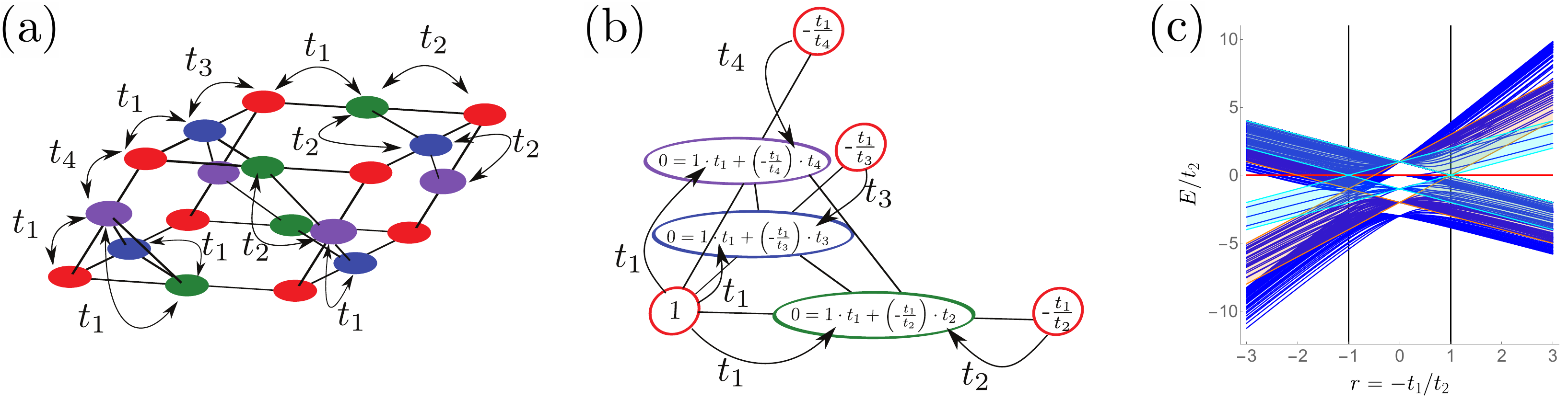}}
\caption{(a) Schematic depiction of the breathing pyrochlore lattice with the $A$, $B$, $B'$, and $B''$ sublattices in red, green, blue, and purple, respectively, and the nearest-neighbor hopping parameters $t_1$, $t_2$, $t_3$, and $t_4$. (b) Illustration of the exact wave function in Eq.~(\ref{eqcornerstatethreedim}) with $r = -t_1/t_2$, $r' = -t_1/t_3$, and $r''= -t_1/t_4$ for a lattice with $M = M' = M''=2$ with the weight of the wave function on each side written inside the red ($A$), green ($B$), blue ($B'$), and purple ($B''$) sites. (c) Energy spectrum with $M = M' = M'' = 5$ and $t_3/t_2 = t_4/t_2 = 1$ such that $r = r' = r'' = -t_1/t_2$ with the bulk bands in blue and the corner mode corresponding to the exact solution in red. The bands in the orange and cyan shaded areas originate from the surface and hinge spectrum, respectively. The black vertical lines indicate where $|r|=1$.}
\label{figpyrochlore}
\end{figure*}
As an illustrative example, we now consider the breathing pyrochlore lattice shown in Fig.~\ref{figpyrochlore}(a), which was also studied in Refs.~\onlinecite{kunstvmiertbergholtz, ezawapap}. This lattice can be thought of as consisting of a stack of breathing kagome lattices that are coupled via intermediate triangular lattices forming tetrahedral structures. The sites in the intermediate triangular lattice play the role of the $B''$ sites. The relevant hopping parameters are shown in Fig.~\ref{figpyrochlore}(a), and we again see that certain $B$, $B'$, and $B''$ sites are connected with each other. The Bloch Hamiltonian for this model reads
\begin{align}
\mathcal{H}_{\vec{k}} &= \begin{pmatrix}
h_A&h_{A,B}&h_{A,B'}&h_{A,B''}\\
h_{A,B}^\dagger&h_B&h_{B,B'}&h_{B,B''}\\
h_{A,B'}^\dagger&h_{B,B'}^\dagger&h_{B'}&h_{B',B''} \\
h^\dagger_{A,B''} & h^\dagger_{B,B''} & h^\dagger_{B',B''} & h_{B''}
\end{pmatrix}, \\
h_{B,B''} &\equiv h_{B,B'';+}+e^{-i (k_3 - k_1)}h_{B,B'';-},\\
h_{B',B''} &\equiv h_{B',B'';+}+e^{-i (k_2 - k_3)}h_{B',B'';-},
\end{align}
which closely resembles Eq.~(\ref{eqblochhamcodimthree}) and $h_{A,B}$, $h_{A,B'}$, $h_{A, B''}$, $h_{B,B'}$ given by Eqs.~(\ref{eqblochhamcodimtwotwotwo}), (\ref{eqblochhamcodimtwotwo}), (\ref{eqblochhamcodimthreetwo}) and (\ref{eqblochhamcodimtwolargertwo}), respectively, with
\begin{align}
& h_A = h_B = h_{B'} = h_{B''} = 0, \nonumber \\
& h_{A,B;+} = h_{A,B';+} = h_{A,B'';+} = h_{B,B';+} \nonumber \\
& = h_{B,B'';+} = h_{B',B'';+} =  - t_1, \nonumber \\
& h_{A,B;-} = h_{B,B';-} = h_{B,B'';-} = h_{B',B'';-} =  - t_2, \nonumber \\
&h_{A,B';-} = - t_3, \qquad h_{A,B'';-} = - t_4. \label{eqhambreathingpyrochlore}
\end{align}
In case of open boundary conditions in all three directions with $A$ sites at its corners, as schematically depicted in Fig.~\ref{figpyrochlore}(a), the Hamiltonian is given by Eq.~(\ref{eqgeneralhampyrcorner}) with $\mathcal{H}_{AB,B';\pm}$ in $\mathcal{H}^{M,M'}$ given by Eqs.~(\ref{eqperphamatobprimetwotwoextra}) and (\ref{eqperphamatobprimetwoextra}), and $\mathcal{H}_{ABB',B'';\pm}$ reading
\begin{widetext}
\begin{equation*}
\mathcal{H}_{ABB',B'';+} =  \begin{pmatrix}
h_{AB,B'';+} & 0 & 0 & 0 \\
h_{B',B'';+} \mathbb{I}_{M \times M} & 0 & 0 & 0 \\
0 & h_{AB,B'';+} & 0 & 0 \\
0 & h_{B',B'';+} \mathbb{I}_{M \times M}& \cdots & 0 \\
0 & \vdots & \ddots & 0\\
0 & 0 & 0 & h_{AB,B'';+}
\end{pmatrix},
\end{equation*} \begin{equation*}
\mathcal{H}_{ABB',B'';-} =  \begin{pmatrix}
h_{AB,B'';-} & 0 & 0 & 0 \\
0& h_{B',B'';-} \mathbb{I}_{M \times M} & 0 & 0 \\
0 & h_{AB,B'';-} & \cdots & 0 \\
0 & \vdots & \ddots & 0\\
0 & 0 & 0 & h_{B',B'';-} \mathbb{I}_{M \times M} \\
0 & 0 & 0 & h_{AB,B'';+}
\end{pmatrix},
\end{equation*}
\end{widetext}
with
\begin{equation*}
h_{AB,B'';+} = \begin{pmatrix}
h_{A,B'';+} & 0 & 0 & 0 \\
h_{B,B'';+}& 0 & 0 & 0 \\
0 & h_{A,B'';+} & 0 & 0 \\
0 & h_{B,B'';+}& \cdots & 0 \\
0 & \vdots & \ddots & 0 \\
0 & 0 & 0 & h_{A,B'';+}
\end{pmatrix},
\end{equation*} \begin{equation*}
h_{AB,B'';-} = \begin{pmatrix}
h_{A,B'';-} & 0 & 0 & 0 \\
0& h_{B,B'';-} & 0 & 0 \\
0 & h_{A,B'';-} & \cdots & 0 \\
0 & \vdots & \ddots & 0 \\
0 & 0 & 0 & h_{B,B'';-} \\
0 & 0 & 0 & h_{A,B'';-}
\end{pmatrix},
\end{equation*}
which differ from Eqs.~(\ref{eqhoppingbetweenbbpprime}) and (\ref{eqhoppingbetweenbbpprimetwo}) by the inclusion of extra hopping terms between $B$ and $B''$ ($B'$ and $B''$) sublattices.

The exact wave function is given in Eq.~(\ref{eqcornerstatethreedim}) and is unaltered by the extra hopping terms as it only depends on the details of the Hamiltonian on sublattice $A$ and the hoppings between the $A$ and $B$ ($B'$) [($B''$)] sites. The terms in the Hamiltonian are given in Eq.~(\ref{eqhambreathingpyrochlore}), such that Eqs.~\eqref{eqgeneralexprforr}, \eqref{eqgeneralexprforrprecise}, and \eqref{eqletslabelanotherone}, which correspond to destructive interference on the $B$, $B'$, and $B''$ sites, respectively, read
\begin{equation}
t_1+r \, t_2=0, \qquad t_1+r' \, t_3=0, \qquad t_1 + r'' \, t_4 = 0,
\end{equation}
where the subscript $1$ on $r$, $r'$, and $r''$ is dropped to simplify the notation, such that $r=-t_1/t_2$, $r'=-t_1/t_3$, and $r''=-t_1/t_4$. The exact wave function is shown explicitly in Fig.~\ref{figpyrochlore}(b) and the energy spectrum in Fig.~\ref{figpyrochlore}(c) with the eigenenergy of the exact solution $h_A=0$ in red. As before, the values of $|r|$, $|r'|$, and $|r''|$ determine where the zero-energy mode localizes; the mode localizes to the corner $\{m,m',m''\} = \{1,1,1\}$, $\{M,1,1\}$, $\{1,M',1\}$, $\{1,1,M''\}$, $\{M,M',1\}$, $\{M,1,M''\}$, $\{1,M',M''\}$, or $\{M,M',M''\}$ when $\{|r|, |r'|, |r''|\} = \{<1, <1, <1\}$, $\{>1, <1, <1\}$, $\{<1, >1, <1\}$, $\{<1, <1, >1\}$, $\{>1, >1, <1\}$, $\{>1, <1, >1\}$, $\{<1, >1, >1\}$, or $\{>1, >1, >1\}$, respectively. The band spectrum in Fig.~\ref{figpyrochlore}(c) is computed for $t_3/t_2 = t_4/t_2 = 1$ such that $r = r' = r''$, and the zero-energy mode localizes to the corner $\{m,m',m''\} = \{1,1,1\}$ ($\{M,M',M''\}$) when $|r|<1$ ($>1$) as was also pointed out in Ref.~\onlinecite{kunstvmiertbergholtz}.

\section{Exactly solvable hinge states on mirror symmetric lattices} \label{secmirrorsymm}

\noindent Lastly, we focus on finding wave functions that localize at mirror-symmetric corners and hinges with codimension two of the type shown in Fig.~\ref{figschemlatmir}. The structures we consider are composed of two different types of one-dimensional chains, which are shown in red and green, and in blue corresponding to a chain made up of zero-dimensional $A$ and $B$ lattices and zero-dimensional $B'$ lattices, respectively. Compared to the structure shown in Fig.~\ref{figschemlat}(b) the main difference is that now the $B'$ chain consists of two sublattices (in Fig.~\ref{figschemlatmir} this is shown with two blue sites instead of only one). Earlier we remarked that it is crucial for the retrieval of exact wave-function solutions that both the $B$ and $B'$ sublattices host a single degree of freedom. There are, however, some instances where this condition can be relaxed; if we couple the $AB$ and $B'$ chains in a mirror-symmetric fashion we again retrieve wave-function solutions of the form of Eq.~(\ref{eqmainresultinsupmat}). Moreover, we again remark that this construction can be easily expanded to higher dimensions as is demonstrated in the explicit examples that follow.

We assume that the $A$ sublattices host $n$ degrees of freedom and $B$ one. Moreover, we refer to the two inequivalent sublattices of the $B'$ chains as $B'_I$ and $B'_{II}$ sites, which are located between $A$ and $B$ sublattices, respectively, and carry $n'_I$ and $n'_{II}$ degrees of freedom, respectively. We require that the $A$ lattices only hybridize with the $B$ and $B'_I$ sites and not with the $B'_{II}$ sites. The corresponding creation operators are denoted by $c^\dagger_{A_i,m,m'}$, $b^\dagger_{m,m'}$, $b'^\dagger_{I,i'_1,m,m'}$, and $b'^\dagger_{II,i'_{II},m,m'}$, respectively, with $i=1,\ldots,n$, $i'_I=1,\ldots,n'_{I}$, and $i'_{II}=1,\ldots,n'_{II}$. We assign to each $B$ site the same unit-cell index as the $A$ site to its left, and to each $B'_I$ ($B'_{II}$) site the same index as the $A$ ($B$) site below. The corresponding Fourier-transformed creation operators are defined as before [see Eqs.~(\ref{eqfouriertransformc})-(\ref{eqfouriertransformctwo})] and we find that the Bloch Hamiltonian is given by $H_{\vec{k}}=\Psi^\dagger_{\vec{k}}\mathcal{H}_{\vec{k}}\Psi_{\vec{k}} $ with  $\Psi^\dagger_{\vec{k}} = (c^\dagger_{A_i,\vec{k}}, b^\dagger_{\vec{k}}, b'^\dagger_{I,\vec{k}}, b'^\dagger_{II,\vec{k}})$ and
\begin{align}
\mathcal{H}_{\vec{k}}&=\begin{pmatrix}
h_A&h_{A,B}&h_{A,B'_I}&0\\
h_{A,B}^\dagger&h_{B}&0&h_{B,B'_{II}}\\
h_{A,B'_I}^\dagger&0&h_{B'_I}&h_{B'_I,B'_{II}}\\
0&h_{B,B'_{II}}^\dagger&h_{B'_I,B'_{II}}^\dagger&h_{B'_{II}}
\end{pmatrix}, \label{eqblochhammirrorsymm}\\
h_{A,B} &\equiv h_{A,B;+}+e^{-ik_1}h_{A,B;-},\\
h_{A,B'_I} &\equiv h_{A,B'_I}+e^{-ik_2}h_{A,B'_I}, \\
h_{B,B'_{II}} &\equiv h_{B,B'_{II}}+e^{-ik_2}h_{B,B'_{II}}, \\
h_{B'_I,B'_{II}} &\equiv h_{B'_I,B'_{II};+}+e^{-ik_1}h_{B'_I,B'_{II};-}. \label{eqblochhammirrorsymmtwo}
\end{align}
In contrast to the notation used so far, we have not included subindices $\pm$ in the $n\times n'_I$ matrix $h_{A,B'_I}$ and the $1 \times n'_{II}$ matrix $h_{B,B'_{II}}$, since the coupling in the $\vec{a}_2$ direction between the $A$ and $B'_I$ ($B$ and $B'_{II}$) sites is symmetric. For simplicity, we have assumed that there is no hopping among the $B$ and $B'_{I}$ lattice sites, but as before including these terms does not alter our findings.

Next, we consider the system with open boundary conditions with $A$ lattices at its corners. Here, we provide the generic Hamiltonian associated with the lattice shown in Fig.~\ref{figschemlatmir}. In the case $M=M'$, which is relevant for the examples that follow, the Hamiltonian is given by $H^{M\times M} = \Psi^\dagger \mathcal{H}^{M,M} \Psi$ with a total of $[M\times M- f(M)]$ $A$ lattices, where $f(M) = 4 \sum_{\alpha = 1}^{(M-1)/2} \alpha$ and $\Psi^\dagger$ is a row vector of creation operators, which reads $
\Psi^\dagger=\left({\bf d}^\dagger_1, \,{\bf e}^\dagger_1, \, {\bf d}^\dagger_2, \,{\bf e}^\dagger_2, \, \ldots, \,{\bf e}^\dagger_{M-1}, \, {\bf d}^\dagger_{M}\right)$ with
\begin{align*}
{\bf d}^\dagger_{m'}&= {\bf a}^\dagger_{1+\lvert m'-\left(M+1\right)/2\rvert,m'},  b^\dagger_{1+\lvert m'-\left(M+1\right)/2\rvert,m'}, \\
& \ldots, b^\dagger_{M-1-\lvert m'-\left(M+1\right)/2\rvert,m'}, {\bf a}^\dagger_{M-\lvert m'-\left(M+1\right)/2\rvert,m'}, \\
{\bf e}^\dagger_{m'}&={\bf b}'^\dagger_{I,3/2+\lvert m'-M/2\rvert,m'},{\bf b}'^\dagger_{II,3/2+\lvert m'-M/2\rvert,m'}, \\
& \ldots, {\bf b}'^\dagger_{II,M-3/2-\lvert m'-M/2\rvert,m'},{\bf b}'^\dagger_{I,M-1/2-\lvert m'-M/2\rvert,m'},
\end{align*}
where ${\bf a}^\dagger_{m,m'}$ is defined in Eq.~(\ref{eqdefvectorstwo}), and ${\bf b}'^\dagger_{j,m,m'}=b'^\dagger_{j,1,m,m'},\ldots,b'^\dagger_{j,n'_j,m,m'}$ with $j=I,II$. Note that the label $m$ of ${\bf a}^\dagger_{m,m'}$, $b^\dagger_{m,m'}$, ${\bf b}'^\dagger_{I,m,m'}$, and ${\bf b}'^\dagger_{II,m,m'}$ explicitly depends on $m'$ due to the varying length of $m$ for different $m'$. The matrix $\mathcal{H}^{M,M}$, which is explicitly different from the matrix defined in Eq.~(\ref{eqgeneralham}), reads
\begin{widetext}
\begin{equation}
\mathcal{H}^{M,M} = \begin{pmatrix}
\mathcal{H}^1_{AB} & \mathcal{H}^1_{AB,B';+} & 0 & 0 & 0 & 0 & 0 & 0 & 0 & 0 \\
\mathcal{H}_{AB,B';+}^{1,\dagger} & \mathcal{H}^1_{B'_IB'_{II}} & \mathcal{H}_{AB,B';-}^{1, \dagger} & 0 & 0 & 0 & 0 & 0 & 0 & 0 \\
0 & \mathcal{H}_{AB,B';-}^{1} & \mathcal{H}^2_{AB} & \cdots & 0 & 0 & 0 & 0 & 0 & 0 \\
0 & 0 & \vdots & \ddots & \mathcal{H}_{AB,B';+}^{\frac{M-1}{2}} & 0 & 0 & 0 & 0 & 0 \\
0 & 0 & 0 & \mathcal{H}_{AB,B';+}^{\frac{M-1}{2},\dagger} & \mathcal{H}^{\frac{M-1}{2}}_{B'_IB'_{II}} & \mathcal{H}_{AB,B';-}^{\frac{M-1}{2}, \dagger} & 0 & 0 & 0 & 0 \\
0 & 0 & 0 & 0 & \mathcal{H}^{\frac{M-1}{2}}_{AB,B';-} & \mathcal{H}^{\frac{M+1}{2}}_{AB} & \mathcal{H}^{\frac{M-1}{2}}_{AB,B';-} & 0 & 0 & 0 \\
0 & 0 & 0 & 0 & 0 & \mathcal{H}_{AB,B';-}^{\frac{M-1}{2},\dagger} & \mathcal{H}^{\frac{M-1}{2}}_{B'_IB'_{II}} & \cdots & 0 & 0 \\
0 & 0 & 0 & 0 & 0 & 0 & \vdots & \ddots & \mathcal{H}_{AB,B';-}^{1} & 0 \\
0 & 0 & 0 & 0 & 0 & 0 & 0 & \mathcal{H}_{AB,B';-}^{1,\dagger} & \mathcal{H}^{1}_{B'_IB'_{II}} & \mathcal{H}_{AB,B';+}^{1,\dagger} \\
0 & 0 & 0 & 0 & 0 & 0 & 0 & 0 & \mathcal{H}^{1}_{AB,B';+} &\mathcal{H}^{1}_{AB}
\end{pmatrix}, \label{eqgeneralhamnew}
\end{equation}
\end{widetext}
\begin{figure*}[t]
  \adjustbox{trim={0\width} {0\height} {0\width} {0\height},clip}
{\includegraphics[width=\linewidth]{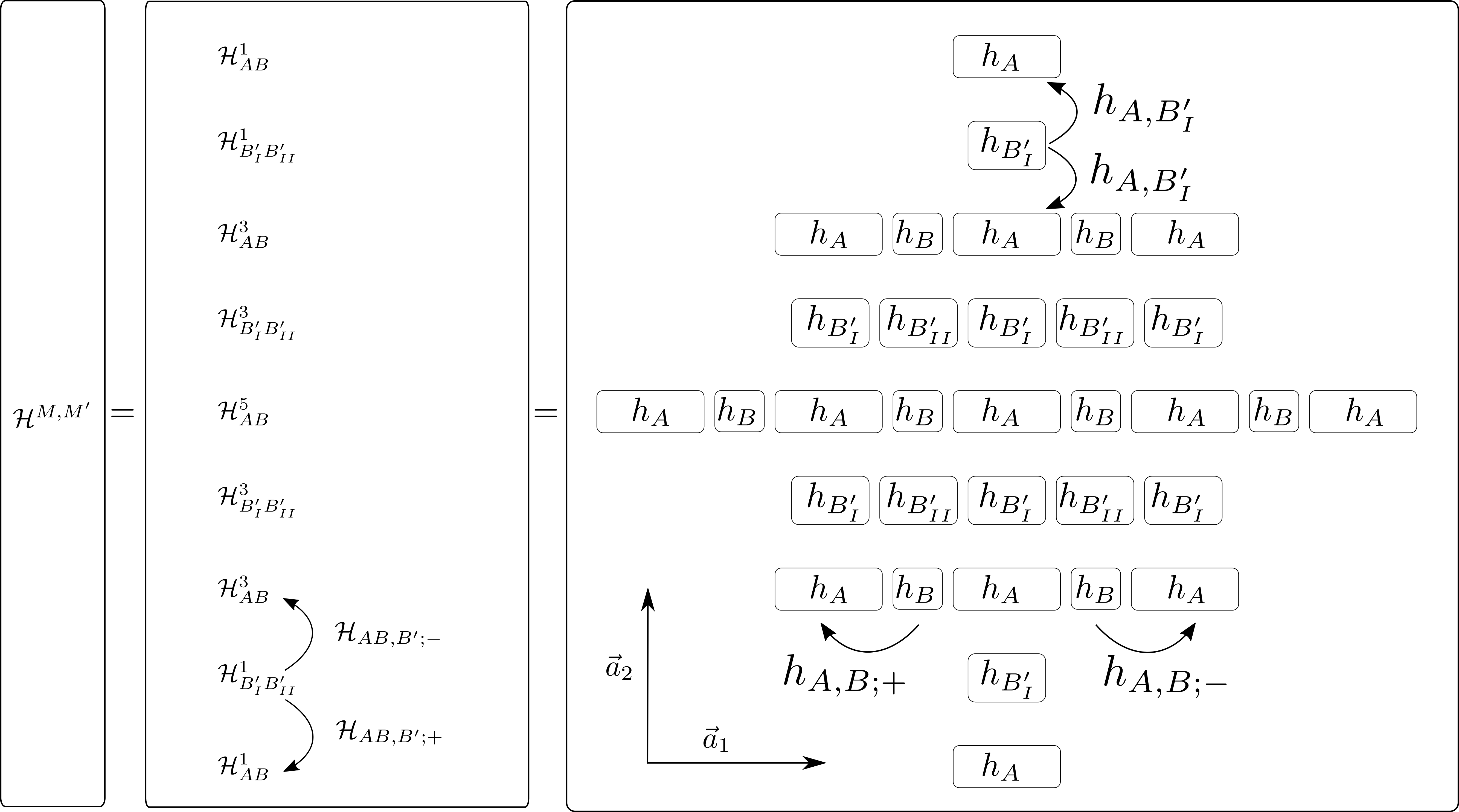}}
\caption{Schematic representation of the Hamiltonian $\mathcal{H}^{M,M}$ in Eq.~(\ref{eqgeneralhamnew}).}
\label{fig2dschemmir}
\end{figure*}
$\mathcal{H}^{m'}_{AB}$ and $\mathcal{H}^{m'}_{B'_IB'_{II}}$ are both given by Eq.~\eqref{eqgeneralhamonerow} with $2m'-1$ $A$ lattices for the first and $h_A\rightarrow h_{B'_I}$, $h_{B}\rightarrow h_{B'_{II}}$, and $h_{A,B;\pm} \rightarrow h_{B'_I, B'_{II};\pm}$ for the latter. The hybridization between the $AB$ and $B'$ chains is captured by $\mathcal{H}^{m'}_{AB,B';\pm}$
\begin{equation}
\mathcal{H}^{m'}_{AB,B';+} = \begin{pmatrix}
h_{A,B'_I} & 0 & 0 & 0 &0&0 \\
0 &h_{B,B'_{II}} & 0 &0 & 0&0 \\
0&0&h_{A,B'_I} & 0 &0 & 0 \\
0&0&0 & h_{B,B'_{II}} &\cdots & 0\\
0 & 0&0&\vdots & \ddots & 0 \\
0 & 0 &0& 0&0 & h_{A,B'_I}
\end{pmatrix},
\end{equation} \begin{equation}
\mathcal{H}^{m'}_{AB,B';-} = \begin{pmatrix}
0 & 0 & 0 & 0 &0&0\\
0 & 0 & 0 & 0&0 &0\\
h_{A,B'_I} & 0 & 0 & 0&0 &0\\
0 &h_{B,B'_{II}}& 0 & 0&0 &0\\
0&0&h_{A,B'_I} & 0 & 0&0\\
0&0&0 &h_{B,B'_{II}} & \cdots & 0\\
0&0&0&\vdots&\ddots & 0 \\
0 & 0 & 0 &0&0& h_{A,B'_I} \\
0 & 0 & 0 &0& 0&0 \\
0 & 0 & 0 & 0&0 &0
\end{pmatrix}. \label{eqperphambnew}
\end{equation}
The full Hamiltonian is shown schematically in Fig.~\ref{fig2dschemmir}. 

To find the exact corner-state wave functions, we need to solve the equations that correspond to destructive interference on the $B$ and $B'_I$ sites. Since the $B'_I$ site may in principle host multiple orbitals, we have to solve $1+n'_I$ equations. Note that as we do not hybridize $A$ lattices with the $B'_{II}$ sites, we get destructive interference for free for the latter. Starting with the ansatz given in Eq.~(\ref{eqmainresultinsupmat}), we obtain the following equations:
\begin{equation}
(h^\dagger_{A,B;+} )_{1,i}+ r_i \, (h^\dagger_{A,B;-})_{1,i}=0, \label{eqevenmorerstwo}
\end{equation} \begin{equation}
(h_{A,B'_I}^{\dagger} )_{l,i}+ r'_i \, (h_{A,B'_I}^{\dagger})_{l,i}= 0, \qquad l=1,\ldots,n'_I. \label{eqevenmorers}
\end{equation}
Note that the second set of equations is trivially solved by $r'_i=-1$, which is due to the mirror symmetry. Therefore, we find that the mode described by the wave function in Eq.~(\ref{eqmainresultinsupmat}) is localized to the $A$ sublattice $\{m,m'\} = \{1,(M+1)/2\}$ and $\{M,(M+1)/2\}$ when $|r_i| <1$ and $>1$, respectively. We point out that the above results also hold when $M \neq M'$.

\begin{figure*}[t]
{\includegraphics[width=\linewidth]{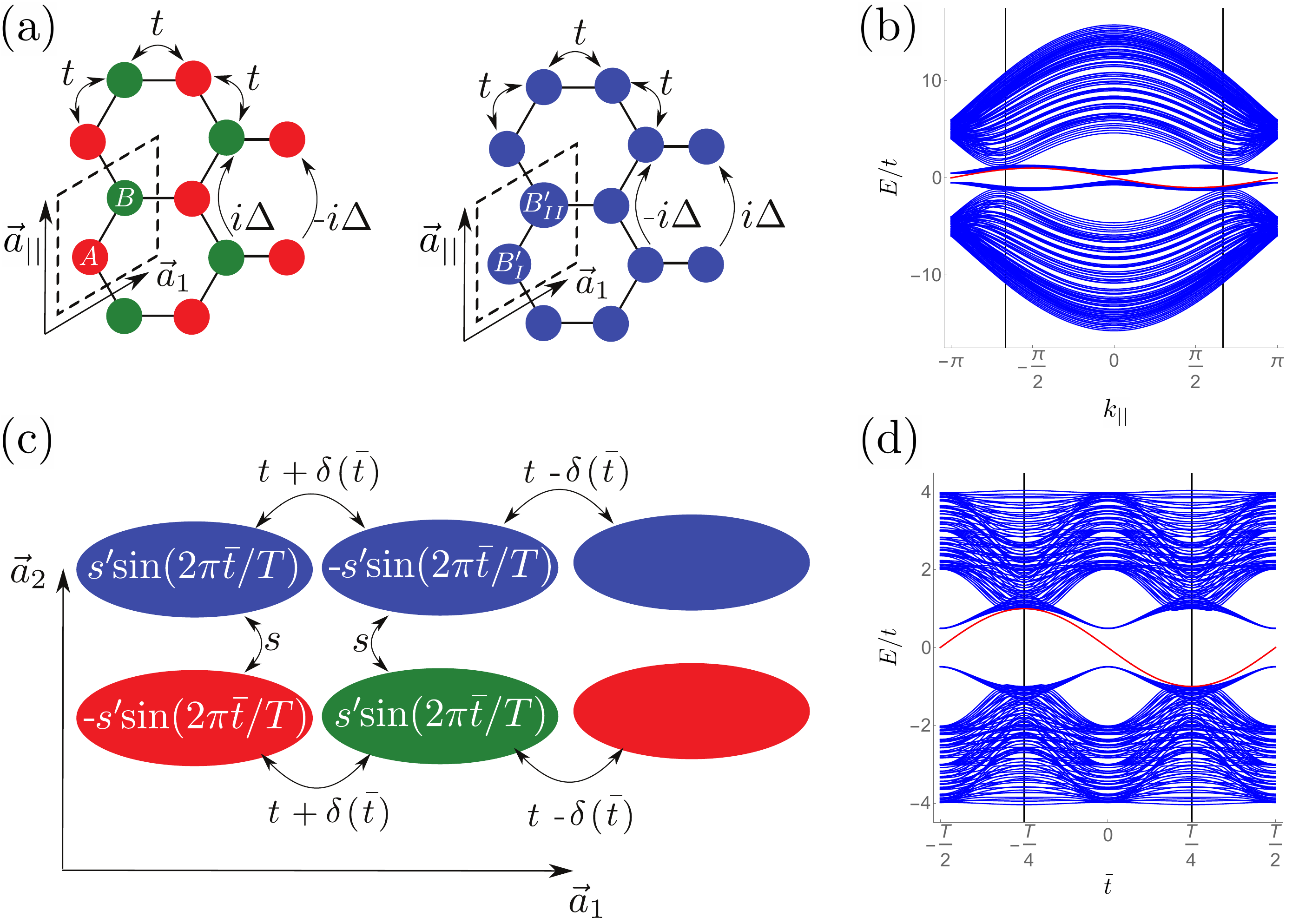}}
\caption{(a) Schematic depiction of the two honeycomb lattices that make up each row $m'$ in Fig.~\ref{figschemlatmir}, with the $A$, $B$, and $B_I$ sublattices in red, green, and blue, respectively, with (next-)nearest neighbor hopping parameters $t$ ($\Delta$) and lattice vectors $\vec{a}_1$ and $\vec{a}_{||}$. Note that the sign of $\Delta$ changes between the two lattices. (b) Energy spectrum of the stacked honeycomb lattices with $M = M' = 11$ and $-10 s/t = 10 \Delta/t = 1$ with the bulk bands in blue and the chiral, hinge mode corresponding to the exact solution in red. The black vertical lines indicate where $|r|=1$. (c) Schematic depiction of the Rice-Mele chains with the same conventions as in panel (a). (d) Energy spectrum of the stacked Rice-Mele chains with $M=M' = 11$ and $3/2 s'/t = -3 s/t = 1$, and the same conventions as in panel (b).}
\label{fighoneyandrm}
\end{figure*}
In Ref.~\onlinecite{kunstvmiertbergholtz}, we showed that we can realize models on this lattice structure that are topologically protected by virtue of the mirror Chern number $C_m (k_2) = [C_+(k_2) - C_-(k_2)]/2$, where $C_{+(-)}(k_2)$ corresponds to the Chern number in the even (odd) sector of the mirror eigenvalues. If $C_m(0) + C_m(\pi) = 1 \, {\rm mod} \, 2$, where $k_2 = 0, \pi$ are the mirror-invariant slices in the Brillouin zone, there are an odd number of chiral, boundary modes at the boundaries that are symmetric under $\vec{a}_2 \rightarrow - \vec{a}_2$. If we thus require that on each $AB$ sublattice we implement a model that has a nonzero Chern number opposite to that of the model on $B'$ sublattices, we find that the boundary modes of neighboring sublattices gap out and only those states on the layer $m' = (M' + 1)/2$ remain. In this case, while the ``ordinary" Chern numbers $C(0)$ and $C(\pi)$ are zero, the mirror Chern number yields the desired situation, i.e., $C_m(0) + C_m(\pi) = 1 \, {\rm mod} \, 2$, implying the presence of boundary modes exponentially localized to $\{m,m'\} = \{1, (M'+1)/2\}$ and $\{M, (M'+1)/2\}$. 

We supplement the above findings with two explicit examples, which were only briefly discussed in Ref.~\onlinecite{kunstvmiertbergholtz}. As a first example, we consider a stack of honeycomb lattices, where each of the $A$, $B$, $B'_I$, and $B'_{II}$ sublattices represents a one-dimensional periodic chain with one degree of freedom such that each of the individual $m'$ $AB$ and $B'$ chains in Fig.~\ref{figschemlatmir} forms a honeycomb lattice with a zigzag edge on one side and a bearded edge on the other as shown in Fig.~\ref{fighoneyandrm}(a). On each honeycomb lattice, we realize a Haldane-like tight-binding model \cite{haldane} as shown in Fig.~\ref{fighoneyandrm}(a) for which we find an exact hinge-state solution. The corresponding Fourier-transformed Hamiltonian is given by Eqs.~(\ref{eqblochhammirrorsymm})-(\ref{eqblochhammirrorsymmtwo}) with
\begin{align}
h_A &= -h_B = -h_{B'_I} = h_{B'_{II}} = -2 \Delta \sin(k_{||}), \nonumber \\
h_{A,B;+} & = h_{B'_I,B'_{II};+} = -t\left(1 + e^{i k_{||}} \right), \label{eqhamhoneycombmirror}\\
h_{A,B;-}& =h_{B'_I,B'_{II};-} = -t , \quad h_{A,B'_I} = h_{B,B'_{II}} = s, \nonumber
\end{align}
where we include hoppings among $B$, $B'_I$, and $B'_{II}$ sites, and $t$ and $s$ (and $\Delta$) are (next-)nearest-neighbor hopping parameters, as shown Fig.~\ref{fighoneyandrm}(a). The term $\pm2 \Delta \sin(k_{||}) $ originates from purely imaginary next-nearest-neighbor hoppings $\pm i\Delta$ in the direction $\vec{a}_{||}$, and realizes a Chern-insulating phase on each individual honeycomb lattice when $\Delta \neq 0$ while an ordinary semimetallic phase is observed for $\Delta = 0$. The opposite sign of $\Delta$ in the $AB$ and $B'$ planes ensures that the two honeycomb lattices have opposite Chern number, such that the two different planes possess counterpropagating edge states. Even though this is a three-dimensional model, we can view it as a collection of two-dimensional models parametrized by $k_{||}$. In particular, we may consider periodic boundary conditions in the $\vec{a}_{||}$ direction and open in the directions $\vec{a}_1$ and $\vec{a}_2$ with the termination shown in Fig.~\ref{fig2dschemmir}. Using the notation introduced above, we have $n=n'_I=n'_{II}=1$, and the Hamiltonian is given in Eqs.~(\ref{eqgeneralhamnew})-(\ref{eqperphambnew}) with the hopping terms in Eq.~(\ref{eqhamhoneycombmirror}). Equations~(\ref{eqevenmorerstwo}) and (\ref{eqevenmorers}) thus read
\begin{equation*}
t\left(1+e^{-ik_{||}}\right) + r(k_{||}) \, t =0, \qquad s+ r' \, s= 0,
\end{equation*}
where we dropped the index $1$ in $r$ and $r'$ for simplicity. This leads to $r(k_{||}) = -\left(1+e^{-ik_{||}}\right)$ and $r' = -1$. The band spectrum is shown in Fig.~\ref{fighoneyandrm}(b), and the chiral hinge state described by Eq.~(\ref{eqmainresultinsupmat}) has eigenenergy $h_A=-2 \Delta \sin(k_{||})$ shown in red.

As a second example, we consider a stack of Rice-Mele chains \cite{ricemele}. Each $A$, $B$, $B'_I$, and $B'_{II}$ sublattice is formed by a single site with one degree of freedom such that we literally consider the lattice as given in Fig.~\ref{figschemlatmir}. We implement the Rice-Mele model on each individual $AB$ and $B'$ chain as shown in Fig.~\ref{fighoneyandrm}(c). The Bloch Hamiltonian is given by Eqs.~(\ref{eqblochhammirrorsymm})-(\ref{eqblochhammirrorsymmtwo}) with
\begin{align}
h_A &= -h_B = -h_{B'_I} = h_{B'_{II}} = s'\sin(2\pi \bar{t}/T), \nonumber \\
h_{A,B;+} & = h_{B'_I,B'_{II};+} = t(1+\tilde{\delta}(\bar{t})), \nonumber \\
h_{A,B;-}& =h_{B'_I,B'_{II};-} = t(1-\tilde{\delta}(\bar{t})), \nonumber \\
h_{A,B'_I}& = h_{B,B'_{II}} = s, \label{eqhamrmmirror}
\end{align}
where $s' = -1$, $t$, and $s$ are nearest-neighbor hopping parameters, $\tilde{\delta}(\bar{t})=\delta(\bar{t})/t=\cos(2\pi \bar{t}/T)/t$, and $T$ is the period of driving as shown in Fig.~\ref{fighoneyandrm}(c). Considering open boundary conditions, we find that the Hamiltonian is given by Eqs.~(\ref{eqgeneralhamnew})-(\ref{eqperphambnew}) with the hopping terms in Eq.~(\ref{eqhamhoneycombmirror}) and with $n = n'_I = n'_{II} = 1$. The band spectrum is shown in Fig.~\ref{fighoneyandrm}(d). This leads to the following for Eqs.~(\ref{eqevenmorerstwo}) and (\ref{eqevenmorers}),
\begin{equation*}
[t+\delta(\bar{t})] + r(\bar{t}) \, [t- \delta(\bar{t})]=0, \qquad s+ r' \, s= 0,
\end{equation*}
where we dropped the index $1$ for simplicity. Hence, at each instant $\bar{t}$ we find precisely one corner mode with energy $s'\sin(2\pi \bar{t}/T)$, or a hinge mode in time, and its associated wave-function solution given in Eq.~(\ref{eqmainresultinsupmat}) with $r(\bar{t}) = - [t + \delta (t)]/[t - \delta (t)]$ and $r' = -1$.

\section{Discussion} \label{secconclusion}

\noindent In this work, we have presented a systematic and coherent method on how to construct lattices on which generic local tight-binding models feature exactly solvable states exponentially localized to boundary motifs of arbitrary codimension and with a dispersion that can be tuned at will by tuning the local hopping amplitudes. By adjusting the hopping parameters that determine the connection of the $A$ sublattices to the $B^{(s)}$ sublattice(s), we have full access to all single-particle properties, such as the $(D-d)$ localization lengths, of these modes. The validity and exact form of our solutions are completely independent of the topological properties of the model as well as the size of the system, and we have demonstrated that while they are no longer exact solutions, they also provide useful insights in the phase diagram of models whose boundaries preserve unit cells, i.e., boundaries made up of $A$ sublattices on one end and $B^{(s)}$ sublattices on the other. While we have restricted ourselves to the dimensions relevant to our three-dimensional physical world in the investigation of different concrete examples, this method can be straightforwardly extend to describe systems of any dimension $D$ with boundaries of any dimension $d<D$. The results presented in this paper were summarized in a shorter previous work in Ref.~\onlinecite{kunstvmiertbergholtz} and have been extended and explored here in more detail.

We point out that our bottom-up approach to engineer lattices with the wished-for properties is extremely versatile and transparent and provides useful insights both in the experimental and theoretical context. In particular, in the latter context, our method shines a rare but clear light on the mechanisms underlying the appearance of boundary modes of any codimension. While several other methods have been introduced to find exact wave-function solutions in systems with open boundaries with codimension one \cite{transfer1, transfer2, transfer3, alasecobaneraortizviola, duncanoehbergvaliente, cobaneraalaseortizviola}, none of these methods have been generalized to describe higher order phases so far, and it is not \emph{a priori} obvious that it is at all possible. Indeed, while all of these methods have different strengths, the clear advantage of our approach is that its generalization to boundaries of any codimension is simple and straightforward. This is also advantageous in the experimental setting in the sense that our method allows for the proposal of simple lattice structures that realize interesting phenomena. The type of lattices that we devise can be straightforwardly realized in artificial lattices such as cold atom systems, optical experiments, microwave cavities, and acoustic metamaterials \cite{mukherjeespracklenvalienteanderssonetc, weimannkremerplotniklumeretc, jotzumesserdesbuquoislebratuehlingergreifesslinger, polibelleckuhlmortessagneschomerus, polischomerusbelleckuhlmortessagne}. This statement is underlined by the recent experimental observation of corner modes in the breathing kagome lattice \cite{xueyanggaochongzhang, elhassankunstmoritzandlerbergholtzbourenanne}.

This work does not explicitly focus on establishing the topological properties of the boundary modes we realize beyond looking at whether the state corresponds to an in-gap state and changes localization upon a band-gap closing, and the validity of the exact solutions is independent of whether the ground state of a model is topological. We, nevertheless, point out that the topological robustness of several of the modes we have realized can be straightforwardly established. Indeed, for models with boundaries of codimension one, topological invariants have been studied in numerous examples, e.g., the end modes of the SSH chain are protected by chiral symmetry, which allows for the definition of a winding number, whereas the chiral modes of a Chern insulator are protected by a nonzero Chern number. We make use of the latter when realizing chiral hinge states in the pyrochlore lattice as well as constructing mirror-symmetric lattices that feature chiral hinge modes, which are protected by the mirror Chern number. To establish the topological robustness of higher order models in general, one may make explicit use of the crystalline symmetries that protect these modes to define invariants \cite{vanmiertortix}, whereas a more generic approach is to use the Wilson-loop formalism to show topological protection \cite{benalcazarbernevighughesagain}. A quick analysis, however, results in the conclusion that the lack of a spectral symmetry, e.g., $E_k = -E_{k}$, in the spectrum of the breathing kagome and pyrochlore models prevents the odd number of corner modes we find at half filling from being topologically protected. Indeed, when introducing a defect on the $A$ sublattice, the zero-energy mode may be pushed away into the bulk, while the localization and exact form of the wave function is unchanged when the defect is introduced on any of the $B^{(s)}$ sublattices. This simple consequence of our construction clearly demonstrates the stability of boundary modes even when they are not explicitly topologically protected; while altering the bulk spectrum, a defect on one or more $B^{(s)}$ sublattice sites does not alter the properties of the boundary mode. This finding is especially relevant in the case of large lattices, i.e., $M^{(s)} \rightarrow \infty$, where the explicit role of the $A$ and $B^{(s)}$ sublattices is washed out away from the boundaries. As a last note, we point out that upon turning of certain couplings such that the breathing kagome and pyrochlore lattices take the form of the Lieb lattice [cf. Fig.~\ref{figschemlat}(b)] and its three-dimensional cousin [cf. Fig.~\ref{figschemlat}(c)], respectively, the spectral symmetry is restored by virtue of which the corner modes in these lattices captured by Eq.~(\ref{eqexactsolgen}) are protected.

The advantage of the transparency of our method has in fact already been proven to bear fruit. Notably, it was very recently demonstrated that topological phases in the context of non-Hermitian systems, which set themselves apart from topological phases in Hermitian models by the general breaking down of conventional bulk-boundary correspondence, can be directly described in a system with open boundary conditions using biorthogonal quantum mechanics \cite{kunstedvardssonetc}. This important insight was strongly inspired by the use of our exact solutions which readily generalize to the non-Hermitian setting. This work thus serves as a supplement to our previous works and can be viewed as a comprehensive guidebook on how to engineer lattices from the bottom up with the desired topological properties up to any order.

\acknowledgments
F.K.K and E.J.B. are supported by the Swedish Research Council (VR) and the Wallenberg Academy Fellows program of the Knut and Alice Wallenberg Foun- dation. G.v.M. is supported by the research program of the Foundation for Fundamental Research on Matter (FOM), which is part of the Netherlands Organization for Scientific Research (NWO).

\end{document}